\documentclass[a4paper,11pt]{article}
\pdfoutput=1 % if your are submitting a pdflatex (i.e. if you have
\usepackage{jheppub} % for details on the use of the package, please
\usepackage[T1]{fontenc} % if needed
\usepackage[utf8]{inputenc}
\usepackage{amsmath,amssymb,amstext,amsfonts} % AMS' mathematical symbols

\usepackage[dvipsnames]{xcolor}
\usepackage{graphicx}
\usepackage{hyperref}
\usepackage{cancel} %for strike-through font
\usepackage{float}
\usepackage{bbm}
\usepackage{mathtools}
\usepackage{lscape}
\usepackage{siunitx}
\usepackage{subcaption}
\usepackage{cleveref}
\usepackage{amsmath}
\usepackage{rotating}
\usepackage{adjustbox}
\usepackage{comment}

\usepackage{slashed}
\usepackage{physics}
\usepackage{nicematrix}

\makeatletter
\renewcommand*\env@matrix[1][\arraystretch]{%
  \edef\arraystretch{#1}%
  \hskip -\arraycolsep
  \let\@ifnextchar\new@ifnextchar
  \array{*\c@MaxMatrixCols c}}
\makeatother

\makeatletter
\g@addto@macro\bfseries{\boldmath}
\makeatother

\definecolor{mygray1}{gray}{0.4}

\def\beq{\begin{equation}}
\def\eeq{\end{equation}}
\def\bsp#1\esp{\begin{split}#1\end{split}}

\usepackage{tikz}
    \usetikzlibrary{positioning}
    \usetikzlibrary{arrows}
    \usetikzlibrary{shapes}
    \usepackage{pgfplots}
    \usetikzlibrary{calc}
    \usetikzlibrary{decorations.markings}
     \usetikzlibrary{decorations.pathmorphing}
    \tikzset{snake it/.style={decorate, decoration=snake}}
\def\centerarc[#1](#2)(#3:#4:#5) % Syntax: [draw options] (center) (initial angle:final angle:radius)
    { \draw[#1] ($(#2)+({#5*cos(#3)},{#5*sin(#3)})$) arc (#3:#4:#5); }
    \usepackage{booktabs}
\usepackage[compat=1.1.0]{tikz-feynman}
\usepackage{contour}

\title{\boldmath 
Integrand Analysis, Leading Singularities and Canonical Bases beyond Polylogarithms}

\author[a]{Felix Forner,} 
\author[b]{Cesare Carlo Mella,}
\author[c]{Christoph Nega,} % \note{Corresponding author.}}
\author[a]{Lorenzo Tancredi,} % \note{Also at Some University.}}
\author[a]{and Fabian J. Wagner}

\affiliation[a]{Technical University of Munich, TUM School of Natural Sciences, Physics Department, James-Franck-Straße 1, D-85748 Garching, Germany}
\affiliation[b]{Rudolf Peierls Centre for Theoretical Physics, University of Oxford,
Clarendon Laboratory, Parks Road, Oxford OX1 3PU}
\affiliation[c]{Max Planck Institute for Gravitational Physics (Albert Einstein Institute), 14476 Potsdam, Germany}

\emailAdd{felix.forner@tum.de}
\emailAdd{cesare.mella@physics.ox.ac.uk}
\emailAdd{christoph.nega@aei.mpg.de}
\emailAdd{lorenzo.tancredi@tum.de}
\emailAdd{fabianjohannes.wagner@tum.de}

\preprint{\begin{minipage}[t]{8cm}\begin{flushright} 
OUTP-26-03P \\ TUM-HEP-1601/26 
      \end{flushright}\end{minipage}}

\abstract{
In this paper, we elaborate on the connection between leading singularities and canonical bases of Feynman integrals beyond polylogarithms.
We start by discussing a notion of leading singularities in dimensional regularization, which can be generalized from the Riemann sphere to more complex geometries, and use it to demonstrate how selecting Feynman integrals with unit leading singularities necessitates introducing new transcendental functions 
related to the periods of the underlying geometries.
Integrals with unit leading singularities in this generalized sense, satisfy $\epsilon$-factorized differential equations, and the new transcendental functions are in direct correspondence to the new differential forms appearing in their Gauss-Manin connection. 
We argue that this construction is mathematically equivalent to the splitting of the period matrix into semi-simple and unipotent parts plus a clean-up step, and demonstrate its use with examples of increasing complexity that require the interplay of multiple geometries. 
}

\begin{document} 
\maketitle
\flushbottom

% !TEX encoding = UTF-8 Unicode
% !TEX root = main.tex
\section{Introduction}
\label{sec:intro}
The calculation of Feynman integrals is central to the exploration of both phenomenological and formal aspects of perturbative quantum field theory. Over the last two decades, in particular, it has become clear that considering suitable bases of Feynman integrals not only can simplify their calculation, but can also help render manifest otherwise hidden structures in physical observables. Despite the lack of a precise mathematical definition of which properties such bases should satisfy in complete generality, it has become common lore to refer to these special choices of Feynman integrals as \emph{canonical bases} or \emph{canonical integrals}~\cite{Arkani-Hamed:2010pyv, Henn:2013pwa}.
The original definition of canonical integrals in physics was largely motivated by the expected simplicity of scattering amplitudes in gauge theories. One of their essential properties is that they should (locally) only exhibit \emph{logarithmic singularities}. 
This property can, in fact, be made manifest \emph{globally}, if all Feynman integrals can be expressed in terms of so-called Multiple Polylogarithms~\cite{Kummer, Goncharov:1995, Remiddi:1999ew, Gehrmann:1999as} (MPLs) or more general iterated integrals over logarithmic differential forms (dlog-forms). Borrowing the terminology from the case where all dlog-forms can be parametrized rationally, we refer to these cases as iterated integrals defined on the Riemann sphere.
A lot is understood about these integrals. In particular, we can associate to every logarithmic integration a so-called \emph{transcendental weight}, which provides the starting point to define a graded (Hopf) algebra and systematize functional relations among these functions~\cite{Goncharov:2010jf, Duhr:2011zq, Duhr:2012fh}. 

For what concerns this paper, there are two important aspects which characterize canonical integrals on the Riemann sphere: First, when evaluated in dimensional regularization~\cite{tHooft:1972tcz, Bollini:1972ui}, they satisfy exceptionally simple systems of differential equations~\cite{Kotikov:1990kg, Remiddi:1997ny, Gehrmann:1999as, Henn:2013pwa}, whose dependence on the dimensional regulator $\epsilon$ is linear and factorized from the kinematic dependence. Moreover, all differential forms 
are differentials of logarithms (dlog-forms). 
Second, integrals satisfying canonical differential equations can be identified by an analysis of the singularities of their \emph{integrands} at $\epsilon=0$. In particular, it is conjectured that integrands with single poles in all variables and whose residues are numerical constants would constitute the right candidates for such canonical integrals. These are typically referred to as integrals with \emph{unit leading singularities}~\cite{Cachazo:2008vp, Arkani-Hamed:2010pyv, Henn:2013pwa, Henn:2020lye}. The procedure of determining canonical integrals by studying the singularities of their integrands is called \emph{integrand analysis}.
If such a basis of integrals is found, the simple form of their differential equations guarantees that, order by order in the dimensional regulator $\epsilon$, they can be expressed as $\mathbb{Q}$-linear combinations of iterated integrals with the same transcendental weight. 
For this reason, these integrals are said to be \emph{pure} and of \emph{uniform transcendental weight}.

One reason why it has been difficult to provide a completely general definition of canonical bases beyond dlog-forms is the richness of the mathematics underlying their formulation. In fact, it is well known that Feynman integrals can be expressed as iterated integrations over differential forms defined on more general classes of complex manifolds than just the Riemann sphere. As of today, numerous examples are known involving elliptic curves, Calabi-Yau manifolds, and higher-genus hypersurfaces and it remains an open question whether these classes of manifolds are enough to exhaust all cases at all loop orders, or more general mathematical structures will eventually appear.

It is natural to expect that a generalization of canonical bases beyond 
polylogarithms should reduce to standard polylogarithmic canonical integrals 
for values of the kinematics for which the non-polylogarithmic geometries degenerate to the Riemann sphere. In the last few years, different approaches have been devised to identify canonical integrals, either tailored to the elliptic case~\cite{Adams:2016xah, Adams:2018yfj, Dlapa:2022wdu, Yang:2025ofz, Chaubey:2025adn, Chen:2025hzq}, on the use of Ansätze~\cite{Adams:2018bsn, Pogel:2022ken, Pogel:2022vat, Pogel:2022yat} or on a direct generalization of integrand analysis based on mixed Hodge theory combined with the property of unipotence~\cite{Gorges:2023zgv, Duhr:2024uid, Duhr:2025lbz,Maggio:2025jel}. 
An alternative approach to obtain differential equations in polynomial form in $\epsilon$, which can then be rotated to an $\epsilon$-factorized form, was also recently proposed in~\cite{e-collaboration:2025frv, Bree:2025tug}.
Most of these approaches have as a central requirement for the definition of a 
canonical basis, the property of $\epsilon$-factorization of the corresponding differential equations. On the other hand, it is important to stress that not any $\epsilon$-factorized system of differential equations is expected to be canonical in the sense above, see for example the discussion in~\cite{Frellesvig:2021hkr,Frellesvig:2023iwr}.
Thanks to these developments, many classes of problems defined on (multiple) elliptic curves~\cite{Jiang:2023jmk, Giroux:2024yxu, Marzucca:2025eak, Becchetti:2025rrz, Becchetti:2025oyb, Coro:2025vgn}, Calabi-Yau geometries~\cite{Duhr:2022pch, Duhr:2023eld, Duhr:2024hjf,  Duhr:2024bzt, Forner:2024ojj, Maggio:2025jel, Duhr:2025kkq, Duhr:2025ouy, Pogel:2025bca, Bern:2025wyd, Klemm:2024wtd, Driesse:2024feo, Driesse:2026qiz, Bern:2026oqp}, and genus-two hypersurfaces~\cite{Duhr:2024uid} have been solved.
Importantly, when successful, most of these methods produce equivalent results 
and have in common the necessity of complementing a suitable initial choice of 
integrals with a set of transformations involving new transcendental functions and 
whose direct interpretation in terms of leading singularities and integrand analysis is not always entirely transparent.

The goal of this paper is to elucidate the non-trivial relations among integrand analysis, leading singularities, and canonical bases on arbitrary geometries. Building upon a generalized definition of leading singularities~\cite{Bourjaily:2020hjv, Bourjaily:2021vyj} and the ideas developed in~\cite{Gorges:2023zgv, Duhr:2025lbz}, we will provide a definition of canonical integrals in terms of integrals with unit leading singularities, elaborating on the subtleties that one encounters when attempting to determine such integrals 
from integrand analysis only. 
We will argue that, contrary to the polylogarithmic case, integrand analysis at strictly $\epsilon=0$ is not sufficient to fully determine canonical integrals on more general geometries. Ultimately, this can be traced back to the structure of the cohomology of these geometries, and, in particular, to the existence of independent differential forms with higher poles. 
This implies that the master integrals which are the representatives of these differential forms, have (multiple) unavoidable transcendental weight drop(s), and an integrand analysis strictly at $\epsilon=0$ cannot fully capture the correct structure of the differential equations at the same transcendental weight for all Feynman integrals. 
The need to correct for this mismatch is at the basis of the extra steps required in the procedure developed in~\cite{Gorges:2023zgv, Duhr:2025lbz}, compared to what is typically sufficient for polylogarithmic cases. 
To be more specific, we will argue that every differential form
in the cohomology with one extra weight drop requires extending integrand analysis by one order in $\epsilon$. Keeping this into account, we will show that all non-trivial transcendental functions appearing in the rotation to a canonical basis can be inferred by analyzing the integrand at the right order in $\epsilon$. 
We note that important information about some of these new functions can be obtained through the study of the intersection matrix in a canonical basis~\cite{Duhr:2024xsy, Duhr:2025xyy}.

This analysis is, at least in principle, possible at the integrand level for those geometries for which a generalization of multiple polylogarithms and the concept of purity are known, as recently demonstrated in the elliptic case~\cite{Yang:2025ofz}.
On the other hand, as originally advocated in~\cite{Gorges:2023zgv, Duhr:2025lbz}, we will demonstrate that by complementing a more standard integrand analysis at exactly $\epsilon=0$, with the information on the differential equations satisfied by the representatives of the cohomology at hand, we can easily generalize these results to arbitrary geometries. Ultimately, we will illustrate how the procedure introduced in~\cite{Gorges:2023zgv, Duhr:2025lbz} produces integrals with unit leading singularities to the correct $\epsilon$-order.
Since the integrals satisfying these equations are naturally expressed 
in terms of iterated integrals with unipotent monodromy~\cite{Brown:2011wfj}, this procedure could be seen
as a starting point to define the corresponding generalization of multiple polylogarithms on an arbitrary geometry.
While we will have in mind applications to Calabi-Yau and higher-genus Riemann surfaces, it will become apparent that this construction is completely general and is based solely on the study of the homology and cohomology of the corresponding problems. Indeed, as illustrated in~\cite{Duhr:2025lbz}, it can in principle be generalized to arbitrary geometries through the language of (mixed) Hodge theory, which is also at the basis of the construction proposed in~\cite{e-collaboration:2025frv, Bree:2025tug}. 

The rest of this paper is organized as follows: First, we summarize some important general concepts and nomenclature used throughout this paper in \cref{sec:gen}. In particular, we recap some aspects of differential forms, periods, transcendental weight, and the closely related concepts of purity and unipotency. Then, in \cref{sec:idea}, we introduce the general idea of this paper, namely, how to generalize the concept of \emph{integrals with unit leading singularities} to Calabi-Yau manifolds and higher-genus Riemann surfaces (\cref{sec:LS}). 
We first focus on the case $\epsilon=0$ in \cref{sec:leadsingeps0} and then consider $\epsilon\neq0$ in \cref{sec:candiff}, where we describe how integrals with unit leading singularities provide 
a canonical basis which fulfills $\epsilon$-factorized differential equations.
As a first pedagogical example, we review the leading singularity analysis in the polylogarithmic case in \cref{sec:poly}. While one can always choose a basis with only single poles, we also illustrate how the analysis can be adapted when starting from integrals with higher poles. This will be the first step towards generalizing the procedure to more general geometries, where higher poles cannot be avoided. 
In~\cref{sec:ell}, we illustrate this on a toy elliptic example, from two different but equivalent points of view: We will show how to find integrals with unit leading singularities either from an integrand analysis, leveraging the knowledge on pure elliptic Multiple Polylogarithms (eMPLs) (see~\cref{subsec:Elliptic Int Analysis}), or by considering the differential equations satisfied by the leading singularities (see~\cref{subsec:EllipticLeadSingDEQs}). 
While the first method relies on the knowledge of the space of functions, we argue that the second one can be more conveniently generalized, and we apply it explicitly to a K3 example in \cref{sec:K3ex}.
After working with simple toy models, in \cref{sec:subtop}, we consider explicit Feynman integral examples beyond the maximal cut up to three loops and illustrate the insights and methodology discussed in the previous sections. We finally draw our conclusions in \cref{sec:con}. In appendix~\ref{app:sunrise}, we complement the integrand analysis for the cubic elliptic model of \cref{subsec:Elliptic Int Analysis} with an analogous analysis for the arguably most famous elliptic Feynman integral, the equal-mass sunrise integral (restricting on its maximal cut for the sake of simplicity).

\section{Generalities and nomenclature}
\label{sec:gen}

Let us start by summarizing some general concepts and introducing the
nomenclature that we will use in the following sections. Many of these concepts have
varying definitions and meanings across the literature, so we find it useful to
clarify how they are to be understood throughout this work. Our primary focus
will be on the underlying intuition, at the expense of mathematical rigor.
Whenever our terminology departs from the standard mathematical usage, it
should be understood as a physics-motivated shorthand tailored to the applications
relevant for the Feynman integral examples studied here.

We are interested in Feynman integrals computed in dimensional regularization,
with the aim of understanding their Laurent expansions near even integer values
of the dimension. In other words, we implicitly always consider
$d=d_0 - 2 \epsilon$, with $d_0=2n$, $n \in \mathbb{N}$, and focus on the
limit $\epsilon \to 0$. In mathematical terms, we will mainly work in standard
cohomology, even if many of the statements below can be rephrased in terms of
twisted cohomology~\cite{Mastrolia:2018uzb, Mizera:2017rqa}. In this limit, Feynman integrals can be expressed in terms
of iterated integrals defined on complex manifolds, prominently Calabi-Yau
manifolds and higher-genus Riemann surfaces. Elliptic curves are a special
case, corresponding to both genus-one Riemann surfaces and complex
one-dimensional Calabi-Yau manifolds, while the simplest case is that of
polylogarithmic Feynman integrals, which are related to the punctured
genus-zero Riemann sphere. Currently, there is no proof that these classes of
geometries are in any way exhaustive, see~\cite{Bargiela:2025vwl} for a
discussion at two-loop order. While in our discussion we will assume to be
dealing with one of these cases, we stress once more that it should be possible to rephrase our statements for arbitrary geometries in terms of mixed Hodge theory~\cite{GriffithsHarris, Vanhove:2014wqa, Duhr:2025lbz}.

\paragraph{Differential forms.}
A first important ingredient that we will need is the cohomology associated
with these geometries, i.e., the types of differential forms that can be
defined on them. Borrowing the terminology commonly used for elliptic curves
(and, admittedly, with some abuse of language), we will refer to differential
forms with no poles as \emph{forms of the first kind} (or holomorphic forms),
to meromorphic differential forms with higher poles but no residues as
\emph{forms of the second kind}, and to all other forms as \emph{forms of the
third kind}~\cite{GriffithsHarris}.
For example, in the case of the punctured Riemann sphere, there are only forms
with single poles (dlog-forms), meaning that all forms are of third kind. On an
elliptic curve, on the other hand, there is always a form of the first kind
and a form of the second kind, where the latter can be chosen as a
derivative of the former. In addition, one can have an arbitrary number of
third-kind forms, corresponding to extra singularities, or punctures, on the
curve.
Increasing the dimensionality of the geometry leads to Calabi-Yau manifolds.
These are equipped with a unique holomorphic differential form, which we still
call a first-kind form. The remaining cohomology is spanned by differential forms with higher
poles but vanishing residues, and we refer to their representatives as
second-kind forms. On the other hand, increasing the genus of a Riemann
surface introduces more holomorphic forms. A genus-$g$ Riemann surface has
$g$ first-kind forms, which we complement by $g$ suitably normalized
second-kind representatives. In both cases, extra punctures on the
corresponding manifolds can give rise to third-kind forms, which
we associate to lower-dimensional manifolds being removed from the
higher-dimensional one, in the same way that a third-kind form on an elliptic
curve or on the Riemann sphere can be associated to removing a point from
them.

\paragraph{Periods and MUM points.}
The homology of a manifold is spanned by a set of independent integration
cycles. By integrating a basis of differential forms over such a set of
independent cycles, one obtains the \emph{periods} of the
manifold.\footnote{In mathematical terms, the periods are defined by the
pairing of cohomology and homology. For a mathematical introduction to periods
in the context of Feynman integrals, see~\cite{Bonisch:2021yfw} and references
therein.} In the following, borrowing once more the nomenclature used for
elliptic curves, we will refer as \emph{periods} only to those obtained by
integrating the form(s) of the first kind, while we will call
\emph{quasi-periods} the ones related to second-kind forms with
higher poles. Periods defined in this sense will play a fundamental role in
the following. Of course, there are no non-trivial periods on the Riemann
sphere, while an elliptic curve has two periods, corresponding to the
integrations of the form of the first kind on a choice of two independent
cycles.

Periods satisfy ordinary higher-order (\emph{Picard-Fuchs}) differential
equations with rational coefficients and regular singularities, whose degree matches the dimension of the cohomology.\footnote{
To be more precise, with cohomology we 
understand the number of  
forms and cycles associated to the integral representation that we use.} 
At each regular singular point, the differential equations admit 
regular solutions as well as algebraically and/or logarithmically divergent ones. If we focus on the logarithmic behavior, a Picard-Fuchs differential equation of order $n$
admits solutions which, locally at a singular point $x=a$, 
can diverge at
most  as $\log^{n-1}(x-a)$. We refer to the
solutions with no logarithmic behaviour 
as \emph{holomorphic solutions} or as \emph{holomorphic
periods}.  Not every regular singular point
admits all possible logarithmic solutions. If the full tower of logarithms is
required, we call this a point of \emph{Maximal Unipotent Monodromy} (MUM
point).
As an example, the periods of an elliptic curve satisfy a second-order differential equation and, after choosing suitable variables, all its singular points are MUM points.
In the elliptic case, for each MUM point, we call the cycle that locally produces
the holomorphic period the $A$-cycle, and the one that yields the
logarithmically divergent period the $B$-cycle. 
For a general geometry of
the type discussed above, locally close to a MUM point, we refer to the cycles
yielding the holomorphic solution(s) as the \emph{holomorphic cycle(s)} and to
the ones yielding the logarithmically divergent one(s) as the
\emph{non-holomorphic cycle(s)}.

\paragraph{Length and transcendental weight.}
The coefficients in the Laurent series in
$\epsilon$ of Feynman integrals can be expressed in terms of (Chen) iterated
integrals~\cite{ChenSymbol}. These integrals are defined over sets of
differential forms which can, in general, contain combinations of
(quasi-)periods of the associated manifolds and other transcendental functions
derived from them.\footnote{See also~\cite{Duhr:2025lbz} for a general
discussion of the transcendental functions needed for a Calabi-Yau
$n$-fold.} We refer to the number of integrations as the \emph{length} of the
iterated integral.

On the Riemann sphere, there are no non-trivial period integrals, and we only
need to consider iterated integrals over differential forms with single poles
(dlog-forms). In this case, it is customary to associate a concept of
\emph{transcendental weight} to the iterated integrals: Every integration over
a single pole increases the transcendental weight by $1$, while integrating
over a function with a higher pole does not increase it. When this happens, we
say that the integral has a \emph{weight drop}. In this way, a logarithm has
transcendental weight $1$, and a multiple polylogarithm of length $n$ has
transcendental weight $n$. A rational function has transcendental weight zero,
since it can be obtained by integrating another rational function with a
higher pole.
We will refer to a function $f(\underline x)$ which is a $\mathbb{Q}(\underline x)$-linear
combination of iterated integrals with the same transcendental weight as a
function of \emph{uniform transcendental weight} or a UT function. Here,
$\mathbb{Q}(\underline x)$ denotes the space of rational functions over
$\mathbb{Q}$ in the variables $\underline x$. For
example, $f(x) = \log(x) + (1+x)\pi$ is a UT function of transcendental weight $1$.
Considering the Laurent expansion of a Feynman integral close to
$\epsilon=0$, we say that it is of uniform transcendental weight 
if the transcendental weight is uniform at every order in
$\epsilon$ and increases by one with every order. This is equivalent to
assigning to $\epsilon$ transcendental weight ``$-1$''~\cite{Henn:2013pwa}.

The generalization of transcendental weight to iterated integrals beyond
polylogarithms remains a matter of discussion. However, close to a MUM point,
the non-trivial geometries degenerate to the Riemann sphere. We can then
assign to the iterated integrals locally the transcendental weight of
the corresponding polylogarithmic expressions. In this way, we may also grade
the periods by their logarithmic behavior: The holomorphic periods have
locally transcendental weight zero, while the logarithmically divergent ones
have increasing transcendental weight up to $(n-1)$. In this sense, the
holomorphic periods play a special role as the natural generalization of
algebraic functions on these classes of geometries.

\paragraph{Purity and unipotency.}
$\mathbb{Q}$-linear combinations of multiple polylogarithms, i.e., combinations
that involve only rational numbers in the coefficients, are said to be
\emph{pure}, since their transcendental weight is decreased by $1$ under
differentiation~\cite{Henn:2013pwa}. Although a widely accepted
generalization of \emph{purity} for arbitrary classes of iterated integrals
relevant to Feynman integrals remains elusive (see also the discussion
in~\cite{Frellesvig:2023iwr}), we adopt the definition
provided in~\cite{Broedel:2018qkq}, which employs the property of
\emph{unipotency}.

A unipotent function satisfies a unipotent differential equation, i.e., a
differential equation without a homogeneous term. For a set of functions
satisfying a system of coupled first-order differential equations, unipotency
is equivalent to the associated differential equation matrix being nilpotent. A
function is then called \emph{pure} if it is unipotent and its total
differential involves only products of pure functions and differential forms with at most
simple poles. The differential forms that we will encounter may contain
transcendental functions related to the (quasi-)periods of the underlying
manifolds. The above definition then applies with the important caveat that,
when analyzing the pole structure near each singular point, we always use the
appropriate local holomorphic solutions for the periods. Note that this concept of unipotency is at the basis of the construction of multiple polylogarithms on arbitrary geometries~\cite{Brown:2011wfj,Broedel:2018qkq}.

As before, for a function to be pure in dimensional regularization, the
coefficient functions at every order in its $\epsilon$-expansion have to be individually pure.

%===========================================
%===========================================
%===========================================
\section{The general idea}
\label{sec:idea}

With these concepts in place, we can summarize the main idea of this paper.
Starting from a generalization of the concept of \emph{leading singularities}
to the type of geometries we are considering, we will argue that the procedure
introduced in~\cite{Gorges:2023zgv, Duhr:2025lbz} produces bases of integrals
satisfying canonical $\epsilon$-factorized differential 
equations precisely by
normalizing their leading singularities to unity, 
in the sense that we now make more precise.

\subsection{Integration Contours and Leading Singularities}
\label{sec:LS}

To begin with, we need to define what a leading singularity is.
We always assume to analyze manifolds iteratively on one-dimensional fibrations. In this case, 
the independent cycles are entirely determined by the holomorphic and the third kind forms, i.e., by enumerating the contours that encircle all single poles and branch cuts of the corresponding integrands.
For a generic (Feynman) integral, deforming the integration contour onto such
cycles produces new objects that encapsulate important information about the
original integral. Since we perform this analysis
iteratively, this turns the problem into studying iterated one-dimensional pairings of contours and forms. 
If the integral is polylogarithmic to all
orders in $\epsilon$, higher poles are not independent and it is enough to
perform this analysis exactly at $\epsilon=0$. One then selects integrals which
at every step only involve integrations over single poles
(dlog-forms)~\cite{Henn:2020lye}. In this case, the elements of the homology
are the cycles around these single poles, and the associated ``maximally
iterated residues'' are called \emph{leading singularities} (LS). If all LS are
constant, the integral is said to have \emph{unit leading singularities}. In
practice, an integral will typically have multiple different leading
singularities, which cannot all be normalized to constants simultaneously. In
this case, one has to subtract a subset of them such that all the remaining
ones can be rendered constant by a single choice of overall normalization.

Beyond the polylogarithmic case, the cohomology and homology groups are more
complicated, but we can define leading singularities by a natural
generalization of the discussion above, see also~\cite{Bourjaily:2020hjv,
Bourjaily:2021vyj} for a general definition 
and application to elliptic cases.
\begin{center}
\framebox[0.88\textwidth]{
\begin{minipage}{0.85\textwidth}
    \emph{Leading singularities are all functions obtained by
iteratively evaluating the integrand close to $\epsilon = 0$ 
on a basis of closed independent contours.}
\end{minipage}
}
\end{center}

For the classes of geometries we are interested in, this procedure yields
cycle-integrals of the differential forms defined on these geometries. 
Note that, according to this notion, 
an integral in dimensional regularization might have different
leading singularities at different orders in $\epsilon$.
With this in mind, we generalize the concept of integrals with unit leading
singularities as follows:
\begin{center}
\framebox[0.88\textwidth]{
\begin{minipage}{0.85\textwidth}
    \emph{A basis of integrals has unit leading singularities if, locally, all
integrals have maximal transcendental weight, 
all corresponding integrands have at most single poles, 
and all leading singularities associated with the
holomorphic cycle(s) and with the cycles encircling the single poles are
constant to order $\epsilon^0$.}
\end{minipage}
}
\end{center}
Let us make some comments on how this definition should be understood.
\begin{enumerate}
    \item If there exists a basis of integrands with at most single poles, it
    is sufficient to perform the analysis directly at $\epsilon=0$, because contributions
    from higher orders in $\epsilon$ are by construction pure in the sense defined above. 
    This is typically the case for polylogarithmic integrals. If, instead, forms with
    higher poles appear, two complications have to be taken into account. First,
    requiring that all integrals have maximal transcendental weight means that
    each weight drop generated by a double pole should be compensated by
    a power of $1/\epsilon$. Second, in order to expose the LS
    associated with these integrals, after rescaling one should either reduce 
    all higher poles to
    single poles via suitable integration-by-parts identities, or 
    expand and analyze the integrand to order $\epsilon^0$. In practice, expanding
    the integrand and computing it on its independent cycles is typically much more difficult.

    \item On geometries beyond the Riemann sphere, evaluating the integrand
    along the non-holomorphic cycles gives rise to transcendental functions
    whose local expansions contain powers of the logarithm of the local
    variable. Normalizing the LS on the holomorphic cycles also fixes the
    normalization of the LS on the non-holomorphic ones. Locally, their leading
    behavior is controlled by pure powers of the logarithm of the local
    variable, possibly plus suppressed lower-weight pieces. 

    \item The different transcendental weights associated with the LS on the non-holomorphic cycles have an important implication for 
    integrals with higher poles in dimensional regularization. 
    While not strictly necessary according to our definition above,
    after the suitable $\epsilon$ rescaling to account for their weight drops,
    the non-holomorphic cycles
    provide the information needed to infer the overall normalization 
    of their LS without having to expand their integrand to $\epsilon^0$. 
    After this normalization, one typically finds extra LS on the
    holomorphic cycle up to order $\epsilon^0$, which have to be subtracted
    to obtain an integral with unit leading singularities.
\end{enumerate}

\subsection{Leading singularities at $\epsilon = 0$}\label{sec:leadsingeps0}

Let us now illustrate these concepts with explicit examples. As anticipated
above, there is an important subtlety hidden in the existence of forms with
higher poles and in their interplay with dimensional regularization. Since LS
are naturally defined in fixed numbers of dimensions, in this subsection we
first focus on what happens exactly at $\epsilon=0$.

Consider a two-dimensional integral of elliptic type
\begin{align}
I_1 = \int^x \frac{\mathrm dx_1}{y(x_1)} \int^{x_1} \frac{\mathrm dx_2}{x_2 - g}\,, \quad y(x) = \sqrt{x(1-x)(1-a\,x)}\,,\quad a,g \in \mathbb{C}\,,
\end{align}
where, for illustrative purposes, we ignore the lower integration boundary and
assume that the integral is convergent. The innermost integration is over a
third-kind dlog-form with a single pole at $x_2=g$, whose residue is given by
the outermost integral. The latter consists of a first-kind form on the
elliptic curve defined by the equation $y^2 = x(1-x)(1-a\,x)$ and admits two
independent integration contours, the $A$- and $B$-cycles. According to the
definition above, there are two leading singularities associated with this
integral,
\begin{equation}
 I_1 \longrightarrow \left\{ \begin{array}{cl}
   A\text{-cycle:}   & \text{LS}\left[I_1\right]_{A} = \varpi_0 \, , \\[1ex]
    B\text{-cycle:}  & \text{LS}\left[I_1\right]_{B} = \varpi_1 \, ,
\end{array} \right.
\label{eq:LSfk0}
\end{equation}
that correspond to the two periods
\begin{align}
\varpi_0  = \oint_A \frac{\mathrm dx_1}{y(x_1)}\,, \qquad \varpi_1  = \oint_B \frac{\mathrm dx_1}{y(x_1)}\,.
\end{align}
In~\cref{eq:LSfk0} we introduced the notation $\text{LS}\left[f(x)\right]_C$
to denote the leading singularity of the function $f(x)$ associated with the
contour $C$. Close to the MUM point $a=0$, the
holomorphic period $\varpi_0$ is the generalization of an algebraic function
and has transcendental weight $0$. The second period $\varpi_1$, instead,
behaves as $\varpi_1 = \varpi_0 \log(a) + p(a)$, where $p(a)$ is holomorphic.
Then, upon dividing $I_1$ by $\varpi_0$, we obtain an integral with unit
leading singularities according to the definition above,
\begin{equation}
 J_1=\frac{1}{\varpi_0} I_1 \longrightarrow \left\{ \begin{array}{cl}
   A\text{-cycle:}   & \text{LS}\left[J_1\right]_{A} = 1 \, , \\[1ex]
    B\text{-cycle:}  & \text{LS}\left[J_1\right]_{B} = \frac{\varpi_1}{\varpi_0} = \tau \sim \log(a) \, ,
\end{array} \right.
\label{eq:LSfk}
\end{equation}
where we use $\sim$ to indicate the local asymptotic behavior. We see that,
after normalizing the LS of this integral on the $A$-cycle, the LS along the $B$-cycle is fixed
and behaves locally as a pure logarithm, i.e., as an object of transcendental
weight $1$.

To span the cohomology of the elliptic curve, we need an integral with a double
pole. As a natural candidate, we can consider the derivative of the integral
above,
\begin{align}
    I_2 = \frac{\partial}{\partial a} \left( \frac{1}{\varpi_0}I_1 \right)\,.
\end{align}
From~\cref{eq:LSfk} we see that its leading singularities are, by construction,
\begin{equation}
I_2 \longrightarrow \left\{ \begin{array}{cl}
   A\text{-cycle:}   & \text{LS}\left[I_2\right]_{A} = 0 \, , \\[1ex]
    B\text{-cycle:}  & \text{LS}\left[I_2\right]_{B} = \partial_a \tau(a) \sim \frac{1}{a} \, ,
\end{array} \right.
\label{eq:LSfk2}
\end{equation}
i.e., they are identically zero on the $A$-cycle and have locally lower
transcendental weight on the $B$-cycle. This means that, if we consider the
two integrals $\{J_1,I_2\}$ together, they are not both of maximal
transcendental weight and, if we were working in dimensional regularization, we
would have to multiply $I_2(x)$ by $\epsilon^{-1}$.

Importantly, as explained above,~\cref{eq:LSfk2} shows that on the
$B$-cycle the LS, while of lower weight, 
is not properly normalized to a constant and suggests instead
to consider
\begin{align}
    J_2 = \frac{1}{\partial_a \tau}\frac{\partial}{\partial a} \left( \frac{1}{\varpi_0}I_1 \right)
    = \frac{\varpi_0^2}{\Delta(a)}\frac{\partial}{\partial a} \left( \frac{1}{\varpi_0}I_1\right)\,,
    \label{eq:SecCan}
\end{align}
where we defined $\Delta(a) = \varpi_0 \varpi_1' - \varpi_0' \varpi_1$. $J_2$
is now properly normalized and has lower transcendental weight compared to
$J_1$. At strictly $\epsilon=0$, there is nothing more we can say, and our
analysis concludes here. 

As a last example, let us consider a modification of the previous integral,
\begin{align}
I_3 = \int^x \frac{\mathrm dx_1}{(x_1-b)y(x_1)} \int^{x_1} \frac{\mathrm dx_2}{x_2 - g}\,, \quad y(x) = \sqrt{x(1-x)(1-a\,x)}\,,\quad a,b,g \in \mathbb{C}\,.
\end{align}
In $I_3$, the innermost integration remains the same as in $I_1$, while the
second has an extra single pole at $x_1 = b$, in addition to the branch cuts
generated by the square root. Therefore, it corresponds to a form of the third
kind on the elliptic curve. In this case, the last integration leads to three
distinct leading singularities: The first two, as before, correspond to the
integrations over the $A$- and $B$-cycles and give rise to complete elliptic
integrals of the third kind. The third is instead the integral around the pole
at $x_1 = b$, yielding an algebraic function, $1/y(b)$. Notice that this
integral has no weight drop compared to $I_1$.

We can easily normalize the leading singularity associated with the residue by
multiplying the integral by $y(b)$, but, differently from the first example,
this step alone does not yield an integral with unit leading singularities. In
fact, integrating around the $A$- and $B$-cycles we obtain non-trivial
transcendental functions, say $G$ and $\tilde G$, respectively. As
before, $G$ is holomorphic and, locally at $a=0$, one can prove that
$\tilde G\sim [G + c] \log(a) + q$, with $q=q(a,b)$ holomorphic and $c$ a
constant that depends on the choice made for the $B$-cycle.
We find
\begin{equation}
y(b)\,I_3  \longrightarrow \left\{ \begin{array}{rll}
   \text{residue:}  &   \text{LS}\left[y(b)\,I_3\right]_{x_1=b} &= 1  \, , \\
   A\text{-cycle:}   &   \text{LS}\left[y(b)\,I_3\right]_{A} &= G = G \, \text{LS}\left[\frac{I_1}{\varpi_0}\right]_{A}  \,  ,   \\[1ex]
    B\text{-cycle:}  &   \text{LS}\left[y(b)\,I_3\right]_{B} &= \tilde G   \sim [G + c] \, \text{LS}\left[\frac{I_1}{\varpi_0}\right]_{B}  \, .
\end{array} \right.   \label{eq:LStk}
\end{equation}
Eq.~\eqref{eq:LStk} showcases that 
the new leading singularity $G$ is proportional to the one of $J_1$,
see~\cref{eq:LSfk}. It is then clear that, in order to obtain an integral with
unit leading singularities, we have to subtract this additional leading singularity
as
\[
J_3 = y(b) I_3 - G J_1 = y(b) I_3 - G \frac{I_1}{\varpi_0}\, ,
\]
such that
\begin{equation}
J_3  \longrightarrow \left\{ \begin{array}{rll}
   \text{residue:} &  \text{LS}\left[J_3\right]_{x=b} &= 1  \, , \\
   A\text{-cycle:}   &  \text{LS}\left[J_3\right]_{A} &=0  \, ,    \\
    B\text{-cycle:}  &  \text{LS}\left[J_3\right]_{B} &\sim c \log(a)  \, ,
\end{array} \right.   \label{eq:LStksub}
\end{equation}
which also removes the non-normalized term in the leading singularity over the
$B$-cycle. The function $G$ just introduced is an example of a new
transcendental function required to fully determine all leading singularities
beyond the polylogarithmic case, as already noted previously for the elliptic
case~\cite{Gorges:2023zgv, Duhr:2025lbz, Chaubey:2025adn}. The same 
argument, in principle, can be generalized to any Calabi-Yau manifold or higher-genus
surface. An integral corresponding to a third-kind form on those geometries can
have, in addition to the leading singularity coming from the residue at a new
pole, additional LS proportional to the respective forms of the first kind.
These contributions must then be removed by subtraction.

In the examples above, we have shown how to construct integrals with unit leading singularities
strictly at $\epsilon=0$. The case of second-kind forms is special, though,
since they have lower weight than the 
corresponding first-kind forms. When we
switch on dimensional regularization, this fact has important
consequences for the $\epsilon$-dependence of the associated differential
equations, and the procedure described above is
not sufficient to obtain $\epsilon$-factorized differential equations. The
reason is the existence of new leading singularities at higher orders in
$\epsilon$. As we will see, the simplest way to expose these leading
singularities is to derive the differential equations that they satisfy. On the
other hand, they can also be exposed directly working at the integrand level if we first turn all
higher poles into single poles with integration by parts. 

\subsection{Leading singularities and $\epsilon$-factorized differential equations}
\label{sec:candiff}
In this subsection, we elaborate on the
connection between differential equations and leading singularities
in dimensional regularization, highlighting 
how the picture changes beyond the strict $\epsilon=0$ limit 
considered above.

Leading singularities close to $\epsilon =0$ provide crucial information  
on the structure of the differential
equations when $\epsilon \to 0$. 
By deforming the integration contour that defines an integral, one can isolate
only part of the differential equations that it satisfies. The most obvious example is a Feynman integral whose integration contour has been deformed to encircle all (a subset of) the poles corresponding to its propagators: This contour defines the maximal (a generalized unitarity) cut. In this case, all subtopologies (or the subset without the cut propagators) have, by construction, no support on that contour and do not contribute. In this
way, one isolates the homogeneous, or part of the, differential equation
system~\cite{Primo:2016ebd,Primo:2017ipr,Frellesvig:2017aai,Bosma:2017ens}. 

The integrand so obtained can itself have further residues and branch cuts,
which define the LS of the maximal cut, or of that unitarity cut. The LS of a
given cut satisfy the differential equations corresponding to that cut. In the
polylogarithmic case, it was conjectured that a Feynman integral with only
single poles and \emph{unit leading singularities} fulfills canonical,
$\epsilon$-factorized differential equations~\cite{Henn:2013pwa}. The goal of
this paper is to show that this picture extends beyond the polylogarithmic case
if we adopt the definition of unit leading singularities proposed
in~\cref{sec:LS}. Indeed, we will argue that the procedure introduced
in~\cite{Gorges:2023zgv, Duhr:2025lbz} produces Feynman integrals with unit
leading singularities according to our definition.
This procedure can be summarized in terms of four main steps:
\begin{itemize}
 \item[1.] Integrand analysis at $\epsilon=0$ is employed to select the right
 integral candidates corresponding to forms of the first and third kind on the
 given geometry.

\item[2.] The representatives of the second-kind forms at $\epsilon=0$ are
chosen as the derivatives of the corresponding first-kind integrals identified
at step 1.

\item[3.] The period matrix at $\epsilon=0$ is split into a semi-simple and
unipotent part, and the basis is rotated by the inverse semi-simple matrix.
This normalizes the leading singularities of the first- and second-kind
integrals, making them unipotent at $\epsilon=0$.

\item[4.] Every integral with $n$ weight drops is rescaled by $1/\epsilon^n$,
and a final clean-up step is required to remove terms that are not fully
$\epsilon$-factorized. This step is typically performed at the level of the
differential equations.
\end{itemize}

Compared to the polylogarithmic case, the appearance of these extra steps may
look unsatisfactory, especially because they involve different orders
in $\epsilon$. Moreover, new transcendental functions are required that do not appear to
have an immediate interpretation within standard integrand analysis. In the
following sections, we will work out explicit cases of increasing complexity
and show that:
\begin{itemize}
    \item[a.] The existence of forms with higher poles in non-polylogarithmic
    problems is what makes the extra steps necessary. This is already true
    in the polylogarithmic case if we choose candidate integrals
    as the derivative of a canonical one, see~\cref{sec:poly}.

    \item[b.] In the polylogarithmic case, no new transcendental functions are
    required because forms with higher poles are not independent.
\end{itemize}

Moreover, beyond polylogarithms, the new transcendental functions required for
a rotation to a canonical basis are nothing but the leading singularities of
the corresponding integrals and are of two types:
\begin{itemize}
    \item[c.] The first type is conceptually the simplest, as it is related to
    forms of the third kind without higher poles. We have already encountered an example
    in the elliptic case in~\cref{eq:LStk}, and we will see
    that their appearance can be predicted by integrand analysis at
    $\epsilon = 0$ also for more general geometries, see~\cref{sec:subtop}.

   \item[d.] The second type can be traced back to the transcendental weight
   drops induced by the second-kind forms and can be interpreted as LS 
   at higher orders in $\epsilon$. While they can be isolated by
   integrand analysis to higher $\epsilon$-orders, we will argue that using
   differential equations to normalize them is equivalent and
   simpler to generalize to other geometries.
\end{itemize}

\section{A polylogarithmic toy model for higher poles}
\label{sec:poly}
It is useful to begin with a simple polylogarithmic toy model. This model allows us to clarify the concepts discussed earlier and isolate the mechanisms that will persist in more complex geometries. Specifically, we focus on the interplay among higher poles, weight drops, and the need to extend the leading-singularity analysis beyond strictly $\epsilon=0$.
We consider the following family of integrals
\begin{align}
    I_n^C(a,\epsilon) &= \int_C \mathrm dx\, \frac{1}{(1-a\, x)^n\,\sqrt{x(1-x)}}
    \left( x(1-x)(1-a\, x)\right)^\epsilon\,, \quad 0<a<1,\, n\in\mathbb Z\, .
    \label{eq:intpol}
\end{align}
In our model, $\epsilon$ plays the role of the dimensional regulator. We could
consider the integral on an arbitrary contour $C$ but, for definiteness, we fix $C = [0,1]$ from now on.
A simple application of integration by parts identities demonstrates that there
are two independent master integrals: $I_0^C(a,\epsilon)$ and
$I_1^C(a,\epsilon)$.

Indeed, inspecting~\cref{eq:intpol} at $\epsilon=0$, we see that the integrand
can develop a pole either at $x=1/a$ or at $x=\infty$, depending on the value
of $n$. Moreover, in both cases, a closed contour integral around the single
branch cut originating from the square root is not independent due to the
global residue theorem. Finally, as long as $n\in\{0,1\}$, all divergences of
the integrand are logarithmic, and we find two leading singularities
corresponding to the two independent cycles around the two poles,
\begin{align}
\text{LS}\left[ \int\frac{\mathrm dx}{\sqrt{x(1-x)}} \right]_{x=\infty} \sim 1 \,, \qquad
    \text{LS}\left[ \int\frac{\mathrm dx}{(1-a\, x)\,\sqrt{x(1-x)}} \right]_{x=1/a} \sim \frac{1}{\sqrt{1-a}} \, .
\end{align}
This provides at once two candidates with unit leading singularities,
\begin{align}
    J_1 = \sqrt{1-a}\,I_{1}^C(a,\epsilon) \,, \qquad
    J_2 = I_0^C(a,\epsilon)\,, \label{eq:canpoly1}
\end{align}
which exhaust the number of independent master integrals.

Indeed, differentiating under the integral sign and using integration by parts, it is easy to demonstrate that
\begin{align}
    \frac\partial{\partial a} \begin{pmatrix}
        J_1 \\ J_2
    \end{pmatrix} =
    \epsilon\, \begin{pmatrix}
     -\frac{3-a}{a(1-a)} & \frac{3  }{a\sqrt{1-a}} \\
    -\frac{1 }{a\sqrt{1-a}} & \frac{1}{a}
    \end{pmatrix}
  \begin{pmatrix}
        J_1 \\ J_2
    \end{pmatrix} \,. \label{eq:epsfact1}
\end{align}
As anticipated in~\cref{sec:LS}, restricting the analysis to $\epsilon=0$ is
sufficient in this case because the cohomology of the problem is spanned by
third-kind forms only, i.e., by dlog-forms.

\subsection*{A candidate with higher poles}
To see how things change when higher poles are involved, it is instructive to
work out the same polylogarithmic example once again, but starting from a basis
that includes an integral with a double pole. 

Let us
pretend that we did not know the second master integral in advance. We then start from the good first candidate $I_1^C(a,\epsilon)$ and
complete the basis by differentiating it with respect to the kinematic variable,
\begin{equation}
\label{eq:integrandDer}
    \partial_a I_1^C =
    (1-\epsilon) \int_C \mathrm dx\, \frac{x}{(1-a x)^2}\, \frac{(x (1-x)(1-a x))^{\epsilon}}{\sqrt{x(1-x)}}\, .
\end{equation}
Clearly, this integrand has a double pole.

To find a canonical basis starting from this candidate, we can leverage our
general definition of integrals with unit leading singularities. In particular,
consider the leading term in $\epsilon$ of~\cref{eq:integrandDer},
\begin{equation}
    \int_C \mathrm dx\, \frac{x}{(1-a x)^2}\, \frac{u^\epsilon}{\sqrt{x(1-x)}}\, , \qquad \text{where} \quad
    u^\epsilon = (x (1-x)(1-a x))^{\epsilon} \, .
\label{eq:defueps}    
\end{equation}
At $\epsilon =0$, i.e., neglecting the ``reduced twist'' $u^\epsilon$, this integrand has a residue at $x=1/a$ which is not properly normalized.
The cycle around this pole was already used to define the candidate $I_1^C$ and in fact, this LS proportional to the one of the original integral can be removed entirely by differentiating the properly normalized first candidate,
\begin{align}
    \partial_a J_1 = \partial_a \left( \sqrt{1-a} \, I_1^C\right) = \int_C \dd x \frac{\sqrt{1-a} \, u^\epsilon}{\sqrt{x(1-x)} (1-a x)^2}\left( \frac{(2-a) x-1}{2(1-a)} - x \,  \epsilon \right) \, . \label{eq:integrandDernorm}
\end{align}
It is easy to check that~\cref{eq:integrandDernorm} at $\epsilon=0$ has
vanishing residue and no pole at infinity. Working
strictly at $\epsilon=0$, no second independent leading singularity is visible.
This is a manifestation of the mechanism discussed in~\cref{sec:idea}: Because the candidate carries a weight drop, the integrand
analysis must be extended by one order in $\epsilon$.

The integrand in~\cref{eq:integrandDernorm} at order $\epsilon^1$ also has a
double pole, which prevents us from seeing its leading singularities directly, neglecting the contribution from $u^\epsilon$.
However, by applying an integration by parts to the $\epsilon^0$ contribution,
keeping the full dependence on $u^\epsilon$, we can move its double pole to
order $\epsilon^1$, such that all double poles cancel at once,
\begin{align}\label{eq:polylog after ibp}
    \partial_a \left( \sqrt{1-a} \, I_1^{C}(a,\epsilon)\right) &=
    \frac{\epsilon}{a\sqrt{1-a}} \int_C \mathrm dx \left(\frac{a-3}{1-a \, x} + 3  \right)\, \frac{u^\epsilon}{\sqrt{x(1-x)}}
     \,.
\end{align}

From~\cref{eq:polylog after ibp}, we observe that while the integration by
parts removed the double pole at $x=1/a$, it has introduced a non-vanishing
residue at infinity. This step highlights the difference
between the polylogarithmic case and more general geometries. In fact, since
the full cohomology of the Riemann sphere is spanned by differential forms with
single poles only, we are always guaranteed to find such an
integration-by-parts relation generated by a rational function, which expresses
an integrand with a double pole in terms of integrands with at most single
poles. On a more general geometry, such as an elliptic curve, differential
forms with double poles can provide independent elements of the cohomology and an attempt to write them as the derivative of something else will force us
to introduce new transcendental functions~\cite{Broedel:2017kkb}.

For the polylogarithmic case at hand, comparing~\cref{eq:polylog after ibp}
with the definition of $I_1^{C}(a,\epsilon)$, it is clear that the two
integrals have the same transcendental weight. Each of the two terms in the
round bracket in~\cref{eq:polylog after ibp} increases the weight by one
through either the single pole at $x=\infty$ or the one at $x=1/a$.
Taking into account the overall prefactor of $\epsilon$, the transcendental
weight of its $\epsilon$-expansion is, however, not maximal and thus not
aligned with the one of $I_1^{C}(a,\epsilon)$. This is a direct consequence of
the weight drop induced by the double pole. To fix this issue, we rescale this
candidate by $1/\epsilon$,
\begin{align}
    \frac{1}{\epsilon}\partial_a \left( \sqrt{1-a} \, I_1^{C}(a,\epsilon)\right) &=
    \frac{1}{a\sqrt{1-a}} \int_C \mathrm dx \left(\frac{a-3}{1-a \, x} + 3  \right)\, \frac{u^\epsilon}{\sqrt{x(1-x)}}
     \,. \label{eq:secmibp3}
\end{align}
As this integral has now only single poles, we can proceed by analyzing its
leading singularities exactly at $\epsilon = 0$.
In particular, the residue at $x=1/a$ corresponds to the first master integral,
$I_1^{C}(a,\epsilon)$, and generates the leading singularity
\begin{align}
    \text{LS} \left[ \frac{1}{a\sqrt{1-a}}\, \int\frac{\mathrm dx}{\sqrt{x(1-x)}} \left( \frac{a-3}{1-a \, x} + 3  \right) \right]_{x=1/a} \sim \frac{a-3}{a(1-a)}\,.
    \label{eq: leading_sing at 1/a}
\end{align}
The second residue, located at $x=\infty$, corresponds instead to the leading
singularity
\begin{align}
    \text{LS} \left[   \frac{1}{a\sqrt{1-a}}\,\int\frac{\mathrm dx}{\sqrt{x(1-x)}} \left( \frac{a-3}{1-a \, x} + 3  \right) \right]_{x=\infty} \sim  \frac{1}{a\sqrt{1-a}}\,. \label{eq:2ndmaster2ndLS}
\end{align}
Due to the two different leading singularities, this integral is not yet a
fully canonical candidate. However, by isolating the second leading singularity
and normalizing it properly, we obtain a final second independent canonical
candidate, which is, in this case, exactly identical to the one obtained
above in~\cref{eq:canpoly1},
\begin{align}
&\frac{a\sqrt{1-a}}{3}\left[\frac{1}{\epsilon}\partial_a \left( \sqrt{1-a}\, I_1^{C}(a,\epsilon)\right) -   \frac{a-3}{a\sqrt{1-a}}I_1^{C}(a,\epsilon)\right] = \int_C \mathrm dx \frac{(x (1-x)(1-a x))^{\epsilon}}{\sqrt{x(1-x)}}  =J_2\,. \label{eq:intanalysis2ndMI}
\end{align}

In summary, we have found a canonical basis only by integrand analysis,
starting from a candidate integral with a double pole, constructed as the
derivative of a starting integral with single poles. Due to the weight drop
induced by the double pole, we had to extend the analysis one order higher
in $\epsilon$ and use an integration-by-parts identity keeping the full
$\epsilon$-dependence of the integrand. In the
polylogarithmic case, this reduction does not introduce any new differential
forms and, once the double pole is removed, the analysis 
closes within
ordinary dlog-forms. As we will see in the next section, this is no longer true for more complicated geometries.

\subsection*{Leading singularities from differential equations}
Let us now connect the analysis above with the method introduced
in~\cite{Gorges:2023zgv, Duhr:2025lbz}, which is based on splitting the period
matrix into semi-simple and unipotent parts. 

Let us consider again the initial basis of integrals
$$I_1^C(a,\epsilon)\,, \quad  \partial_a I_1^C(a,\epsilon)\, .$$
It is easy to demonstrate that they satisfy the differential equations
\begin{align}
    \frac\partial{\partial a}
    \begin{pmatrix}
        I_1^C \\ \partial_a I_1^{C}
    \end{pmatrix}
    =
    \begin{pmatrix}
       0 & 1 \\
 \frac{1}{2 a(1-a)}+\frac{\epsilon}{2 a(1-a) }-\frac{\epsilon ^2}{a(1-a)} & -\frac{2-5 a}{2 a(1-a) }-\frac{2 \epsilon }{a(1-a) }
    \end{pmatrix}
    \begin{pmatrix}
        I_1^C \\ \partial_a I_1^{C}
    \end{pmatrix}
    \,. \label{eq:eqder1}
\end{align}
This already amounts to steps 1 and 2 of the general method summarized
in~\cref{sec:candiff}. We then proceed with step 3 and look at the
differential equations in~\cref{eq:eqder1} at $\epsilon=0$. This is a coupled
$2\times 2$ system and its complete solution is a $2\times 2$ matrix $W(a)$,
the period matrix, which fulfills
\begin{align}
\label{eq:DEWronskian}
    \frac\partial{\partial a} W(a) =
    \begin{pmatrix}
       0 & 1 \\
\frac{1}{2 a(1-a)} & -\frac{2-5 a}{2 a(1-a) }
    \end{pmatrix} W(a)\,.
\end{align}
Integrals satisfying canonical differential equations are pure functions, which
implies that they are also unipotent, order-by-order in $\epsilon$.
Mathematically,~\cref{eq:DEWronskian} shows that the period matrix $W(a)$ is
\emph{not unipotent}, because it does not satisfy a unipotent differential
equation. In fact, by solving the differential equation, we find
\begin{align}
    W(a) = \begin{pmatrix}
        \varpi_0(a) & \varpi_1(a) \\ \varpi_0'(a) & \varpi_0'(a)
    \end{pmatrix}
    = \begin{pmatrix}
        \frac{1}{\sqrt{1-a}} & \quad & \frac{1}{\sqrt{1-a}}\ln{\left(\frac{1-\sqrt{1-a}}{1+\sqrt{1-a}}\right)} \\ \frac{1}{2 (1-a)^{3/2}} & \quad & \frac{1}{a (1-a)} + \frac{1}{2 (1-a)^{3/2}}\ln{\left(\frac{1-\sqrt{1-a}}{1+\sqrt{1-a}}\right)}
    \end{pmatrix} \, .
\label{eq:pmpoly}
\end{align}
The algebraic prefactors spoil unipotence. Further, we find a sum of terms with
different transcendental weights in the bottom-right entry.

To remove the non-unipotent part of the period matrix, which is usually
referred to as the \emph{semi-simple} part, we identify the unipotent objects
in~\cref{eq:pmpoly}. Explicitly, these are given by the logarithms and
numerical constants, while the algebraic (and rational) functions are not.
This implies the decomposition
\begin{align}
    W(a) = \underbrace{\begin{pmatrix}
         \frac{1}{\sqrt{1-a}} & 0 \\
 \frac{1}{2  (1-a)^{3/2}} & \frac{1}{a (1-a) }
    \end{pmatrix}}_{W_\text{ss}} \cdot
    \underbrace{\begin{pmatrix}
         1 & \ln{\left(\frac{1-\sqrt{1-a}}{1+\sqrt{1-a}}\right)} \\
        0 & 1
    \end{pmatrix}}_{W_\text{u}} \label{eq:WSWUsplit}\, .
\end{align}
Defining a new basis whose period matrix is $W_\text{u}$ is equivalent to
rotating our starting basis by the inverse of the semi-simple part of the
period matrix,
\begin{align}
    \begin{pmatrix}
        J_1 \\ \widetilde{J}_2
    \end{pmatrix} = D_\epsilon\, W_\text{ss}^{-1} \begin{pmatrix}
        I_1^C(a,\epsilon) \\ \partial_a I_1^{C}(a,\epsilon)
    \end{pmatrix} =
    \begin{pmatrix}
        1 & 0 \\ 0 & \frac{1}{\epsilon}
    \end{pmatrix}
    \begin{pmatrix}
        \sqrt{1-a} & 0 \\
 -\frac{a}{2} & a(1-a)
    \end{pmatrix}\begin{pmatrix}
        I_1^C(a,\epsilon) \\ \partial_a I_1^{C}(a,\epsilon)
    \end{pmatrix} \,. \label{eq:ssrotpoly}
\end{align}
Note that we have also rescaled the second master integral by $1/\epsilon$
through the matrix $D_\epsilon$. The necessity of this rescaling can be seen
directly from the form of the unipotent matrix in~\cref{eq:WSWUsplit}: After
rotating by the inverse of the semi-simple part, the newly defined second
master integral contributes with a single pole to the derivative of the first
one,
\begin{align}
   \frac\partial{\partial a} W_\text{u} = \begin{pmatrix}
       0 & \partial_a \ln{\left(\frac{1-\sqrt{1-a}}{1+\sqrt{1-a}}\right)}  \\
       0 & 0
   \end{pmatrix} W_\text{u} \, .
\end{align}
Since the first master has uniform transcendental weight, this equation can
only be consistent if the second master has a lower transcendental weight,
which motivates the $\epsilon$ rescaling.

Let us pause here to compare the rotation we have performed so far with the one
worked out through integrand analysis in~\cref{eq:intanalysis2ndMI}, which
corresponds to
\begin{align}
    \begin{pmatrix}
        J_1 \\ J_2
    \end{pmatrix} &=
    \begin{pmatrix}
        \sqrt{1-a} & 0 \\
 \frac{3-a}{3} & \frac{a\sqrt{1-a}}{3\epsilon}
    \end{pmatrix}\begin{pmatrix}
        I_1^C(a,\epsilon) \\ \partial_a \left( \sqrt{1-a}\, I_1^{C}(a,\epsilon)\right)
    \end{pmatrix}  \nonumber \\
    &=
    \begin{pmatrix}
        \sqrt{1-a} & 0 \\
    \frac{3-a}{3} - \frac{a}{6\epsilon} & \frac{a(1-a)}{3\epsilon}
    \end{pmatrix}
    \begin{pmatrix}
        I_1^C(a,\epsilon) \\ \partial_a I_1^{C}(a,\epsilon)
    \end{pmatrix}\,. \label{eq:canbaspolIA}
\end{align}
Notice that the rotation by the inverse semi-simple matrix
in~\cref{eq:ssrotpoly}, up to an irrelevant overall factor $1/3$ for the second master, already
captures almost completely the full rotation from~\cref{eq:canbaspolIA}. The
only part missing is a contribution at higher order in $\epsilon$
proportional to the first master. In the previous subsection, this piece was
found by analyzing the integrand of the second master one order higher in
$\epsilon$. In the procedure outlined in~\cref{sec:candiff}, it would be dealt with in step 4, utilizing the information coming from the differential equations.

Indeed, the unipotent-semi-simple splitting has been performed exactly at
$\epsilon=0$, see~\cref{eq:pmpoly}. Thus, rotating away the inverse
semi-simple part only takes care of the non-zero leading singularities of the
two masters at their first non-trivial order in $\epsilon$. Due to the weight
drop in the second master, these do not correspond to the same transcendental
weight and therefore are not enough to fully determine the canonical
candidates. In the polylogarithmic case, the missing information can still be
extracted entirely within the space of dlog-forms, in agreement with the
integrand analysis above. In the next section, the same mechanism will reappear
for an elliptic problem, but will require genuinely new
transcendental objects.

\section{The elliptic case: A cubic model}
\label{sec:ell}
In this section, we consider an example closely
related to the one discussed abstractly in~\cref{sec:leadsingeps0}, working this time
in dimensional regularization.
The geometry is still elliptic, but once $\epsilon$ is switched on, 
the weight drop in the second candidate can be compensated by $\epsilon$, 
and its leading singularities must be analyzed up to order $\epsilon^0$.

We modify the family of integrals in~\cref{eq:intpol} and consider instead
\begin{align}
    I_n^C(a,\epsilon) &= 
    \frac{1}{2\sqrt{\pi}} 
    \int_C \mathrm dx\, \frac{x^n}{\sqrt{x(1-x)(1-a\, x)}}
    \left( x(1-x)(1-a\, x)\right)^\epsilon\,, \; 0<a<1\,, \; n \in \mathbb{Z} \, ,  \label{eq:intell}
\end{align}
where the factor $2\sqrt{\pi}$ is a convenient normalization which we can largely ignore in the following. 
As in the previous example, the integrals depend
explicitly on the contour $C$. The equation
\begin{equation}
y^2 = x(1-x)(1-a\, x)\,,
\end{equation}
defines a family of elliptic curves parametrized by the variable $a$.
Moreover, the integrand for $\epsilon=0$ never develops a single pole at
$x=\infty$ for any $n \in \mathbb{Z}$, which means that there is no extra
puncture at infinity and the dimension of the corresponding (co-)homology is
two.

We take the two contours corresponding to the $A$- and $B$-cycles of the elliptic curve  to construct the two periods of the elliptic curve given by
\begin{align}
    \varpi_0 = I_0^A(a,0) &=
    \frac{1}{2\sqrt{\pi}}
    \oint_A \mathrm dx\, \frac{1}{\sqrt{x(1-x)(1-a\, x)}} = 
    \frac{1}{\sqrt{\pi}}
    \int_0^1 \mathrm dx\, \frac{1}{\sqrt{x(1-x)(1-a\, x)}}\,,
    \nonumber \\
    \varpi_1 = I_0^B(a,0) &=
    \frac{1}{2\sqrt{\pi}}
    \oint_B \mathrm dx\, \frac{1}{\sqrt{x(1-x)(1-a\, x)}} = 
    \frac{1}{\sqrt{\pi}}
    \int_1^{1/a} \mathrm dx\, \frac{1}{\sqrt{x(x-1)(1-a\, x)}} \,. \label{eq:periodsell}
\end{align}
Note that, compared to~\cref{eq:intell}, 
in~\cref{eq:periodsell} we assume that whenever we integrate over the $B$-cycle we 
change 
the sign of the root to guarantee that all integrals are real.
With this choice the Legendre relation becomes
\begin{align}
     \Delta(a) = \varpi_0 \varpi_1' - \varpi_1 \varpi_0'
    = \frac{1}{a(a-1)}\,.
\label{legendrerel}
\end{align}
If the integration contour $C$ corresponds to one of the two cycles, it is
easy to see that
\begin{align}
    \oint_{A,B} \mathrm dx\ \frac\partial{\partial x} \left( x^n\, \sqrt{x(1-x)(1-a\, x)}
    \left( x(1-x)(1-a\, x)\right)^\epsilon \right) = 0 \, ,
\end{align}
which implies the recursion relation
\begin{align}
   \frac{1}{2} a (3  + 2 n + 6 \epsilon) I_{n+2}^{A,B}(a,\epsilon)
   - (1+a) (1  + n + 2 \epsilon) I_{n+1}^{A,B}(a,\epsilon)
   + \left( \frac{1}{2} + n + \epsilon \right) I_{n}^{A,B}(a,\epsilon) = 0 \,.\label{eq:recursion}
\end{align}
Clearly, a very similar recursion relation can be derived on an arbitrary
contour $C$, with the obvious difference that, in general, one would require a
boundary term, which in the language of Feynman integrals would correspond to a
subtopology. From~\cref{eq:recursion}, we then see that all integrals can be
reduced to two master integrals, $I_0^C(a,\epsilon)$ and $I_1^C(a,\epsilon)$,
modulo possible boundary terms, as expected from the dimension of the
cohomology. From now on, we ignore boundary terms and drop the superscript $C$
or, in physics language, we work on the maximal cut. This is not an
oversimplification here, since the new phenomenon that we want to describe
has its origin in the structure of the leading singularities of the two
master integrals on the maximal cut.

The recursion relation in~\cref{eq:recursion} allows us to derive the system of
differential equations satisfied by the two master integrals,
\begin{align}
    \frac\partial{\partial a} \begin{pmatrix}
        I_0(a,\epsilon) \\
        I_1(a,\epsilon)
    \end{pmatrix}
    =
    \left(
\begin{array}{cc}
 \frac{1}{2 (1-a)}+\frac{\epsilon }{1-a} & -\frac{1}{2 (1-a)}-\frac{3 \epsilon }{1-a} \\
 \frac{1}{2 a(1-a) }+\frac{\epsilon }{a(1-a) } & -\frac{2-a}{2 a(1-a)}-\frac{(2+a) \epsilon }{a(1-a)} \\
\end{array}
\right)
    \begin{pmatrix}
        I_0(a,\epsilon) \\
        I_1(a,\epsilon)
    \end{pmatrix} \, .
    \label{eq:diffeq1}
\end{align}
This basis is clearly non-canonical: The differential equations are linear in
$\epsilon$ but their complexity is hidden in the fact that they are maximally
coupled at $\epsilon=0$.

Since the geometry of this problem is an elliptic curve without any
extra punctures, the first step to find a canonical basis is to identify a master integral that maps to the form of the first kind at
$\epsilon=0$ and has unit leading singularities. We have already constructed
the right candidate in the example discussed in~\cref{sec:leadsingeps0},
see~\cref{eq:LSfk}, which in the case at hand gives
\begin{equation}
    J_1 = \frac{I_0(a,\epsilon)}{\varpi_0}\,.
\end{equation}

The genuinely new feature appears for the second candidate once dimensional
regularization is switched on. As in~\cref{sec:leadsingeps0}, the natural starting point
is obtained by differentiating the normalized first master. Since this object
has a weight drop, however, it must now be divided by $\epsilon$. We therefore
consider
\begin{equation}
    J_2 =\frac{1}{\epsilon} \frac{\varpi_0^2}{\Delta} \frac\partial{\partial a} \left( \frac{I_0(a,\epsilon)}{\varpi_0}\right) = \frac{\varpi_0 \partial_a I_0(a,\epsilon)
    - I_0(a,\epsilon) \partial_a \varpi_0}{\epsilon\, \Delta(a)} \, .
\end{equation}
Thus, compared to the discussion in~\cref{sec:leadsingeps0}, the only change so far is
the compensating factor of $1/\epsilon$, which keeps track of the weight drop.
Using~\cref{legendrerel}, we can easily verify that its leading singularities
at the first non-trivial order in $\epsilon$ are indeed constant,
\begin{align}
     J_2 &
    \longrightarrow \left\{
    \begin{array}{ccc}
         \text{A cycle:} &  \quad \text{LS}[J_2]_A = \frac{1}{\epsilon} \frac{\varpi_0 \varpi_0' - \varpi_0 \varpi_0'}{\Delta(a)}   + \mathcal{O}(\epsilon^0) = &  \mathcal{O}(\epsilon^0) \, , \hfill \\
         \text{B cycle:} & \quad \text{LS}[J_2]_B = \frac{1}{\epsilon} \frac{\varpi_0 \varpi_1' - \varpi_1 \varpi_0'}{\Delta(a)}   + \mathcal{O}(\epsilon^0) = & \frac{1}{\epsilon} + \mathcal{O}(\epsilon^0) \, .
    \end{array}
     \right.
     \label{eq: leading sing second ell}
\end{align}
The second candidate now starts one order earlier on the $B$-cycle, while the $A$-cycle remains trivial at that order. 
This fixes the overall
normalization of the second master, but not yet its full canonical form.
Putting everything together, this provides the basis
\begin{align}
   \begin{pmatrix}
       J_1 \\ J_2
   \end{pmatrix} =
    \begin{pmatrix}
        1 & 0 \\ 0 & \frac{1}{\epsilon}
    \end{pmatrix}
    \begin{pmatrix}
        \frac{1}{\varpi_0} & 0 \\
       - \frac{\varpi_0'}{\Delta(a)} & \frac{\varpi_0}{ \Delta(a)}
    \end{pmatrix} \begin{pmatrix}
        I_0(a,\epsilon) \\
        \partial_a I_0(a,\epsilon)
    \end{pmatrix}\,. \label{eq:rot1ell}
\end{align}
As in the polylogarithmic case,~\cref{eq:rot1ell} corresponds to a rotation by
the inverse of the semi-simple part of the period matrix, followed by a
rescaling by $1/\epsilon$ of the second master. In fact, in this case, the
period matrix can be written as
\begin{align}
    W(a) = \begin{pmatrix}
        \varpi_0 & \varpi_1 \\
        \varpi_0' & \varpi_1'
    \end{pmatrix} = \underbrace{\begin{pmatrix}
      \varpi_0 & 0 \\ \varpi_0' & \frac{\Delta(a)}{\varpi_0}
    \end{pmatrix}}_{W_\text{ss}}
    \underbrace{\begin{pmatrix}
        1 & \tau \\ 0 & 1
    \end{pmatrix}}_{W_\text{u}} \qquad\text{with}\qquad W_\text{ss}^{-1} = \begin{pmatrix}
    \frac{1}{\varpi_0} & 0 \\
       - \frac{\varpi_0'}{\Delta(a)} & \frac{\varpi_0}{ \Delta(a)}
\end{pmatrix}\,. \label{eq:diffJ1J2ell}
\end{align}
The differential equations for this basis then read
\begin{align}
\frac\partial{\partial a} \begin{pmatrix}
       J_1 \\ J_2
   \end{pmatrix}
&=
   \left[  \epsilon \, \begin{pmatrix}
        0 & \frac{1}{a(a-1)\varpi_0^2} \\
 \varpi_0^2
 & \frac{2  }{a(a-1) }
   \end{pmatrix} +
\begin{pmatrix}
    0 & 0 \\
    2 \varpi_0' \varpi_0 & 0
\end{pmatrix} \right]
\begin{pmatrix}
       J_1 \\ J_2
   \end{pmatrix} \,. \label{eq:diffeq3}
\end{align}
The equations are not yet $\epsilon$-factorized. The violating term in~\cref{eq:diffeq3} 
is precisely the manifestation of an extra leading
singularity associated with the second master, which only shows up at order
$\epsilon^0$. The rest of this section is devoted to 
predicting this missing term in two equivalent ways.

\allowdisplaybreaks
%======================================
%======================================
%======================================
\subsection{Normalizing leading singularities with integrand analysis}\label{subsec:Elliptic Int Analysis}

As a first possibility, we will demonstrate that the extra leading singularity
spoiling the $\epsilon$-factorization of~\cref{eq:diffeq3} can indeed be
identified by an integrand analysis extended to one order higher in $\epsilon$.
This analysis can conveniently be performed referring to the integration
kernels that define elliptic Multiple Polylogarithms (eMPLs)
~\cite{Brown:2011wfj,Broedel:2017kkb,Broedel:2018qkq}, similarly to what was
done in~\cite{Yang:2025ofz}. The conceptual point is that, unlike in the
polylogarithmic warm-up, after the compensating $1/\epsilon$ rescaling the
higher-pole candidate no longer reduces to ordinary dlog-forms, but instead
produces a genuine second-kind elliptic object.

To write our integrands in terms of these kernels, we follow~\cite{Broedel:2017kkb} and define the objects
\begin{equation}
    \begin{aligned}
        c_3&=\frac{\sqrt{a_3-a_1}}{2} = \frac{1}{2\sqrt{a}}\,, &
    \omega_0(a) &= 2c_3\int\limits_{a_1}^{a_2}\frac{\dd x}{\bar{y}} = \int\limits_{0}^{1}\frac{\dd x}{y}\,, \\
    \widetilde\Phi_3(x) &= \frac{1}{c_3 \bar{y}}\left(-x + \frac{s_1(\vec a)}{3}\right) = \frac{2a}{y}\left(-x + \frac{1+a}{3a}\right)\,, \ &
    \eta_0 &= \frac{1}{4}\int\limits_{a_1}^{a_2}\dd x \, \widetilde\Phi_3(x)\,,  \\
    \Phi_3(x) &=\widetilde\Phi_3(x) - 8c_3 \frac{\eta_0}{\omega_0 \bar{y}}=\widetilde\Phi_3(x) - 4 \frac{\eta_0}{\omega_0}\frac{1}{y}\,, &
    Z_3(x) &= \int\limits_{a_3}^x \dd x' \, \Phi_3(x') \,,
    \end{aligned}
\end{equation}
where $\bar{y}=y/\sqrt{a}$ is of the form
\begin{align}
    \bar{y}=\sqrt{(x-a_1)(x-a_2)(x-a_3)} \qquad \text{with} \quad a_1=0 \, , \ a_2=1 \, , \ a_3=1/a \, .
\end{align}
Note that, differently from~\cite{Broedel:2017kkb}, we label the first (quasi-)period with the subscripts $0$ rather than $1$, and $y$ in \cite{Broedel:2017kkb} corresponds to $\bar y$ here. Furthermore, the period $\omega_0$ defined above is related to the period $\varpi_0$ in our convention (see \cref{eq:periodsell}) by $\omega_0= \sqrt{\pi} \varpi_0$.

Expressing our integrand in terms of these quantities allows us to map them to pure eMPL kernels~\cite{Brown:2011wfj,Broedel:2018qkq}, 
which generalize dlogs on the elliptic curve. 
The logic behind this is that, if we succeed to write our integrand entirely in terms of pure eMPL kernels, we are guaranteed that this candidate respects our definition of unit leading singularities without having to explicitly evaluate the LS on the independent cycles. 
This is the natural extension of what happens with standard MPLs and dlog-forms and, in technical terms, it boils down to the fact that eMPLs are multivalued functions with 
unipotent monodromy on the elliptic curve~\cite{Brown:2011wfj}.

We then  aim to write our integrand in terms of the pure kernels $g^{(n)}(z\pm z_i)$ defined on the torus, see~\cite{Broedel:2018qkq}. For a special point $x=c$ on the elliptic curve, we will denote the corresponding point on the torus by $z_c$ in the following.
Let us start by fixing the first integral. 
In the language of pure elliptic
polylogarithms, we immediately see that the holomorphic integral
$I_0(a,\epsilon)$ is canonical up to a normalization by $\varpi_0$, since 
\begin{align}
    I_0(a,0) \propto \int_C \frac{\mathrm dx}{y} \propto \varpi_0\int_C \mathrm d z \, ,   
\end{align}
where we work modulo irrelevant overall numerical prefactors.
We thus choose the first canonical integral as
$J_1 = \frac{I_0(a,\epsilon)}{\varpi_0}$.

For the second integral, we again start from the derivative of the first
canonical integral, $J_1$. Since it has a weight drop due to the double pole, we divide by $\epsilon$ and
keep the full $\epsilon$-dependent expression. We can easily express 
the integrand in terms of the quantities defined above, 
\begin{align}\label{eq:J2ellipptic}
    \frac{1}{\epsilon}\partial_a J_1
    &\propto
    \frac{1}{\epsilon}\frac{\Delta(a)}{\varpi_0}\int_C \dd x \left( \frac{\eta_0}{\sqrt{\pi}\varpi_0y}-\frac{1-2a}{6y}-\frac{(1+6\epsilon)\widetilde \Phi_3(x)}{4}+\frac{1-2a+6\epsilon}{6y}\right)u^\epsilon \nonumber\\
    &=-
    \frac{1}{\epsilon}\frac{\Delta(a)}{\varpi_0}\int_C \dd x \frac{\Phi_3(x)}{4}u^\epsilon
    +  \frac{\Delta}{\varpi_0}
    \left(\int_C \dd x \left( -\frac{3\widetilde \Phi_3(x)}{2}+\frac{1}{y}\right)u^\epsilon\right) \, ,    
\end{align}
where $\Delta(a)$ and $u^\epsilon$ were defined in~\cref{legendrerel}
and~\cref{eq:defueps}, respectively.
Again, this integral contains a double pole, this time isolated
in the functions $\Phi_3(x)$ and $\widetilde \Phi_3(x)$.

It is instructive to first look at the 
$1/\epsilon$-piece, setting $u^\epsilon=1$. This
contribution vanishes on the $A$-cycle, 
which is analogous to what we found in
the polylogarithmic example. 
As opposed to the polylogarithmic case, however,
we can now also consider the $B$-cycle, 
where we find a constant multiple of
$\Delta/\varpi_0^2$. Indeed, this information can be used to 
fix the normalization of the integral to be
$\frac{\varpi_0^2}{\Delta}$. This is of course consistent 
to what was found by our previous analysis
in~\cref{eq: leading sing second ell}. 
As discussed in~\cref{sec:LS}, while we can use the
$B$-cycle to fix the normalization of the second integral, it does not yet tell
us anything about its leading singularities one order higher in $\epsilon$.

The information about the overall normalization
of the second integral is also contained in the $A$-cycle, once we extend its
analysis one order higher in $\epsilon$. To illustrate this, let us ignore for
the moment the information from the $B$-cycle and go back to the derivative of
$J_1$ in~\cref{eq:J2ellipptic} on the $A$-cycle, considering now the next order
in $\epsilon$. Emulating our analysis from the polylogarithmic case, we integrate by parts the
$1/\epsilon$ term, where keeping the reduced twist $u^\epsilon$ 
is essential in order to avoid
boundary terms. We find, again modulo numerical overall prefactors
\begin{align}\label{eq:integrand J2 after IBP}
    \frac{1}{\epsilon}\partial_a J_1 &\propto \frac{\Delta(a)}{\varpi_0}\int_C \dd x \left(\frac{Z_3(x)}{x}+\frac{Z_3(x)}{x-1}+\frac{ Z_3(x)}{ x-1/a} - 6\widetilde \Phi_3(x)\right) u^\epsilon
    +  \Delta(a)\, J_1\,.
\end{align}
While entirely equivalent, at variance with~\cite{Yang:2025ofz}, we perform
integration by parts to match our integrand directly to the eMPL kernels, in
the same fashion as we did in the polylogarithmic case in~\cref{sec:poly}.
Eq.~\eqref{eq:integrand J2 after IBP} makes the
elliptic analogue of the polylogarithmic mechanism explicit: After the
$1/\epsilon$ rescaling the higher-pole candidate 
is rewritten in terms of a primitive $Z_3(x)$ of the second-kind form $\Phi_3(x)$.
At this point, we can finally set $\epsilon=0$ and attempt 
to write our integrand in terms of
pure eMPL kernels.

We first focus on the part in brackets in
\cref{eq:integrand J2 after IBP}. Using
\begin{equation}
    \begin{aligned}
        \frac{Z_3(x)\dd x}{x-c}- 2 \widetilde \Phi_3(x) \dd x& = 4 \frac{\dd z }{\omega_0}\bigg( g^{(2)}(z-z_c)+g^{(2)}(z+z_c)-2g^{(2)}(z)\\
    &\phantom{=}+2 g^{(2)}(z_c)+g^{(1)}(z_c)\big(g^{(1)}(z-z_c)-g^{(1)}(z+z_c)\big)\bigg)\, , \\
    g^{(2)}(z_c) &= \frac{1}{4}Z_3^{(2)}(c) = \frac{1}{2} g^{(1)}(z_c)^2 - \frac{1}{8c_3^2}\left(x-\frac{s_1}{3}\right) \, , \\
    -2 g^{(1)}(z_{a_i}) \dd z &= \big(g^{(1)}(z-z_{a_i})-g^{(1)}(z+z_{a_i})\big) \dd z \,,
    \end{aligned}
\end{equation}
\begin{equation}
    \sum\limits_{i=1,2,3}\left(a_i-\frac{s_1}{3}\right)=0 \, ,
\end{equation}
and the special values
\begin{align}
    g^{(1)}(z_0)=g^{(1)}(z_1)= -\pi i\,, \qquad g^{(1)}(z_{1/a})=0\,,
\end{align}
which can be derived based on~\cite{Broedel:2017kkb}, we find
\begin{align}
     &\frac{1}{4}\left(\frac{Z_3(x)}{x}+\frac{Z_3(x)}{x-1}+\frac{Z_3(x)}{x-1/a} - 6\widetilde \Phi_3(x)\right)\dd x\\
     &\propto \frac{\dd z}{\varpi_0} \bigg( g^{(2)}(z-z_0)+g^{(2)}(z+z_0)+
     g^{(2)}(z-z_1)+g^{(2)}(z+z_1)\\
     & +g^{(2)}(z-z_{1/a})+g^{(2)}(z+z_{1/a})
     -6g^{(2)}(z) - 2\pi^2\bigg)\, .
\end{align}
This proves that the combination
\begin{align}
    \frac{\varpi_0}{4}\left(\frac{Z_3(x)}{x}+\frac{Z_3(x)}{x-1}+\frac{Z_3(x)}{x-1/a} - 6\widetilde \Phi_3(x)\right)\,,
\end{align}
corresponds to a constant linear combination of $g$-kernels on the torus, and
is thus pure.  Comparing with our integrand in
\cref{eq:integrand J2 after IBP}, we see that we need to rescale with
$\varpi_0^2/\Delta(a)$ and then subtract the extra term
$\varpi_0^2(a) J_1$. In fact, choosing
\begin{align}
    K_2 = \frac{1}{\epsilon}\frac{\varpi_0^2}{\Delta(a)}\partial_a J_1 - \varpi_0^2(a)\, J_1 \,,
\label{eq:cleanupell1}
\end{align}
the resulting differential equations
\begin{align}
    \frac\partial{\partial a} \begin{pmatrix}
        J_1 \\ K_2
    \end{pmatrix} = \epsilon
    \begin{pmatrix}
        -\frac{1}{a(1-a)} & -\frac{1}{a(1-a)\varpi_0^2} \\ -\frac{(1-a+a^2)\varpi_0^2}{a(1-a)} & -\frac{1}{a(1-a)}
    \end{pmatrix}
    \begin{pmatrix}
        J_1 \\ K_2
    \end{pmatrix}
\end{align}
are $\epsilon$-factorized.

An important comment is in order. 
At first glance, the result obtained from~\cref{eq:integrand J2 after IBP} might look inconsistent with our statement in~\cref{eq: leading sing second ell}, where we had shown that $J_2$ should have a LS on the $B$-cycle equal to $1/\epsilon$. 
This is not the case. The point is that the integral~\cref{eq:integrand J2 after IBP} is strictly speaking divergent at $\epsilon=0$ only on the $B$-cycle. 
This can be seen by comparing with~\cref{eq:J2ellipptic}, before the final integration by parts. 
The primitive $Z_3(x)$ is defined from $\Phi_3(x)$, whose $A$-period vanishes by construction while its $B$-period does not. 
As a consequence, the singularities at $x=\{0,1,1/a\}$ are regulated by $Z_3(x)$ on the holomorphic cycle, but not on the $B$-cycle. 

This analysis demonstrates that, in the elliptic case, using pure eMPLs, an
integrand analysis can be performed in a similar fashion as when starting from
integrals with higher poles in the polylogarithmic case~\cite{Yang:2025ofz}. 
The key difference is
that the reduction of the higher-pole candidate no longer closes within
ordinary dlog-forms, but produces a genuine elliptic second-kind object, $Z_3(x)$. See
appendix~\ref{app:sunrise} for an analogous treatment of the maximal cut of the equal-mass sunrise graph.
Importantly, this  analysis \emph{relied on the knowledge of the function
space in order to identify pure forms}. Our goal in the next subsection is to
obtain the same information in an equivalent way, which can be generalized to
settings where this knowledge is not available.

%======================================
%======================================
%======================================
\subsection{Normalizing leading singularities from differential equations}\label{subsec:EllipticLeadSingDEQs}

We now show that the shift in~\cref{eq:cleanupell1} can be determined
without referring explicitly to the pure eMPL kernels. 
Instead, we derive
differential equations directly for the leading singularities. This way of
arguing is entirely equivalent to the integrand analysis above, but can be more
easily generalized to cases where the relevant function space is not known.

Let us go back to~\cref{eq:diffJ1J2ell}. By specializing the differential
equations for the two independent contours and using~\cref{eq:LSfk,eq: leading sing second ell}, we find for the first master
\begin{align}
\frac {\partial}{\partial a} \text{LS}[J_1]_{A,B} &= \frac{\epsilon}{a(a-1)\varpi_0^2} \text{LS}[J_2]_{A,B}
 \nonumber \\ &
   = \left\{  \begin{array}{ccl}
         A\text{-cycle:} & \quad & \mathcal{O}(\epsilon )\,, \\
         & \\
         B\text{-cycle:} & \quad & \frac{1}{a(a-1)\varpi_0^2} + \mathcal{O}(\epsilon ) = \frac{\partial}{\partial a} \tau + \mathcal{O}(\epsilon ) \, ,
    \end{array} \right.
\end{align}
as expected. In the second line, we expanded in $\epsilon$ and included the
explicit results for the leading singularities already normalized at the
previous step. On the other hand, for the second master at the same order, we
find
\begin{align}
    \frac {\partial}{\partial a} \text{LS}[J_2]_{A,B} &= 2 \varpi_0' \varpi_0 \text{LS}[J_1]_{A,B} + \epsilon \frac{2}{a(a-1)} \text{LS}[J_2]_{A,B} + \mathcal{O}(\epsilon)
    \nonumber \\ &
   = \left\{  \begin{array}{ccl}
         A\text{-cycle:} & \quad & 2 \varpi_0' \varpi_0 + \mathcal{O}(\epsilon)\,,\\
         &  \\
         B\text{-cycle:} & \quad & 2 \varpi_0' \varpi_0\,  \tau + \frac{2}{a(a-1)} + \mathcal{O}(\epsilon)\,,
    \end{array} \right.
\end{align}
where again we have expanded in $\epsilon$ and included the results for the
leading singularities identified before.

From these equations, we see that the second
master possesses at order $\epsilon^0$ a LS proportional to
that of the first master. It is this extra leading singularity that is behind
the incomplete decoupling of the two masters at order $\epsilon^0$. To remove
it, we focus on the $A$-cycle and shift the second master as
\begin{align}
    K_2 = J_2 - \varpi_0^2(a)\, J_1 = \frac{1}{\epsilon} \frac{1}{\Delta(a)}\left( \varpi_0 \partial_a I_0(a,\epsilon)
    - I_0(a,\epsilon) \partial_a \varpi_0 \right) - \varpi_0^2(a) J_1\,.
\label{eq:cleanupell2}
\end{align}
Indeed, repeating now our analysis for the newly defined master integral $K_2$
we find
\begin{align}
    \frac {\partial}{\partial a} \text{LS}[ K_2]_{A,B}
    &=  \epsilon \frac{2}{a(a-1)} \text{LS}[ J_2]_{A,B} + \mathcal{O}(\epsilon)
   = \left\{  \begin{array}{ccl}
         A\text{-cycle:} & \quad & \mathcal{O}(\epsilon)\,,\\
         & \\
         B\text{-cycle:} & \quad &  \frac{2}{a(a-1)} + \mathcal{O}(\epsilon)\,.
    \end{array} \right.
\end{align}
In other words, up to additive constants,
\begin{align}
  K_2
    \longrightarrow \left\{  \begin{array}{ccl}
         A\text{-cycle:} & \text{LS}[K_2]_A = & \mathcal{O}(\epsilon)\,,\\
         & \\
         B\text{-cycle:} &  \text{LS}[K_2]_B = &  2 \log\left( \frac{1-a}{a} \right)+ \mathcal{O}(\epsilon)\,.
    \end{array} \right.
\end{align}
The leading singularities of the second master are now properly normalized on
both cycles according to the definition given in~\cref{sec:LS}. The result
is, of course, identical to the one inferred by integrand analysis, as is easy
to see by comparing~\cref{eq:cleanupell1,eq:cleanupell2}.

\subsubsection*{Leading singularities and multiple polylogarithms on arbitrary geometries}
Before generalizing our analysis to more complicated cases, let us pause to discuss what we have achieved. In our second approach, we have derived differential equations for the leading singularities of the corresponding Feynman integrals. 
This effectively required being able to perform integration by parts  
to map all integrands to a basis of the 
cohomology of the problem under study. 
These manipulations are indeed equivalent to the ones required to fix leading singularities by integrand analysis to higher $\epsilon$-orders in~\cref{subsec:Elliptic Int Analysis}. 
Also in that case we had to perform integration by parts to map the integrands 
to the ones of pure elliptic multiple polylogarithms, as also discussed in~\cite{Yang:2025ofz}.
On the other hand, the approach based on differential equations 
is more general, as it does not rely on an a-priori knowledge of 
the differential forms defining multiple polylogarithms on arbitrary manifolds. 
From this perspective, differential equations may be viewed not only as a tool for studying Feynman integrals, but also as the starting point for organizing and, possibly, defining the corresponding class of generalized polylogarithmic functions.

\section{Beyond elliptic geometries: A K3 example}
\label{sec:K3ex}
We now apply the same strategy to a geometry beyond the elliptic case.
The point of this section is to show that the construction of unit leading singularities continues to work when no explicit analogue of pure elliptic multiple polylogarithms is directly available.

For the sake of simplicity and to be as explicit as possible, we consider a one-parameter K3 example, which is the natural generalization of the elliptic problem considered above,
\begin{align}
    I_{n,m}^C(a,\epsilon) = \int_C \mathrm dx \mathrm dy \,\frac{x^n y^m }{ \sqrt{x(1-x) y (1-y) (1- a\, x y)}} (x(1-x) y (1-y) (1- a\, x y))^\epsilon\,,
\end{align}
where  $n,m\in\mathbb Z$.
Again, by a simple application of integration by parts identities, we can demonstrate that on any contour $C$ there are three master integrals, modulo possible boundary terms, which we ignore for the present treatment. In particular, the masters can be chosen to be
$I_{0,0}(a,\epsilon)$, $I_{0,1}(a,\epsilon)$, $I_{0,2}(a,\epsilon)$.
It is then easy to see that the first integral has no poles, and it corresponds to the form of the first kind on the K3 geometry. The remaining two have higher poles and never single poles, which makes them good candidates to span the cohomology of the corresponding K3 surface at $\epsilon = 0$.

We can then derive a system of differential equations satisfied by these integrals,
\begin{align}
    \frac\partial{\partial a} \begin{pmatrix}
        I_{0,0}(a,\epsilon) \\
        I_{0,1}(a,\epsilon) \\
        I_{0,2}(a,\epsilon)
    \end{pmatrix} 
    = 
    \begin{pmatrix}
         -\frac{2-a}{2 a(1-a) }+\frac{\epsilon }{1-a} & \frac{4+5 a}{2 a(1-a) }-\frac{\epsilon }{1-a} &     -\frac{9}{2 (1-a)}-\frac{3 \epsilon }{1-a} \\
        -\frac{1}{2 a(1-a) }+\frac{\epsilon }{a(1-a) } & 
        \frac{2+7 a}{2 a(1-a) }-\frac{(2-a) \epsilon}{a(1-a)} & 
        -\frac{9}{2 (1-a)}-\frac{3\epsilon }{1-a} \\
        -\frac{1}{2 a(1-a) }+\frac{\epsilon }{a(1-a) } & \frac{5+4 a}{2 a(1-a) }-\frac{\epsilon }{a(1-a) } & 
        -\frac{4+5 a}{2a (1-a)}-\frac{(2+a) \epsilon}{a(1-a)}
    \end{pmatrix}
    \begin{pmatrix}
        I_{0,0}(a,\epsilon) \\
        I_{0,1}(a,\epsilon) \\
        I_{0,2}(a,\epsilon)
    \end{pmatrix} \, .
\end{align}
While the above equations are linear in $\epsilon$, they are maximally coupled at $\epsilon = 0$.
To go to an $\epsilon$-factorized basis, we first derive the differential equations satisfied by the derivatives of the first master integral, which will constitute important building blocks for our canonical basis. Going to the derivative basis, we find
\begin{align}
\frac\partial{\partial a} \begin{pmatrix}
        I_{0,0} \\
        I_{0,0}' \\
        I_{0,0}''
    \end{pmatrix} 
    = 
    \left( A_0 + \epsilon A_1 + \epsilon^2 A_2 + \epsilon^3 A_3 \right)
    \begin{pmatrix}
        I_{0,0} \\
        I_{0,0}' \\
        I_{0,0}''
    \end{pmatrix} \, ,
\end{align}
where
\begin{equation}
\begin{aligned}
    A_0 &= \begin{pmatrix}
        0 & 1 & 0 \\
        0 & 0 & 1 \\
        \frac{1}{8a^2(1-a)} & \frac{4-13 a}{4 a^2 (a-1)} & \frac{6-9 a}{2 a(a-1) }
    \end{pmatrix}\,, \quad &
      A_1 &= \begin{pmatrix}
        0 & 0 & 0 \\
        0 & 0 & 0 \\
        \frac{1}{4 a^2(1-a)} & \frac{2 (2-a)}{a^2(a-1)} & \frac{4-a}{a(a-1) } 
    \end{pmatrix} \, ,  \\[1ex]
     A_2 &= \begin{pmatrix}
        0 & 0 & 0 \\
        0 & 0 & 0 \\
        \frac{1}{2 a^2(a-1) } & \frac{4+a}{a^2(a-1)} & 0 
    \end{pmatrix}\,, \quad &
    A_3 &= \begin{pmatrix}
        0 & 0 & 0 \\
        0 & 0 & 0 \\
        \frac{1}{a^2(a-1) } & 0 & 0
    \end{pmatrix} \, .
\end{aligned}
\end{equation}
From here on, we suppress the dependencies of $I_{0,0}$ on $a$ and $\epsilon$, and use $I_{0,0}'=\partial_a I_{0,0}$.

From the  formulas above, we can read off the third-order Picard-Fuchs operator which annihilates the first master at $\epsilon = 0$,
\begin{align}
   \left[  \frac{\partial^3}{\partial a^3} +\frac{(6-9 a) }{2 a(1-a) } \frac{\partial^2}{\partial a^2} + \frac{(4-13 a) }{4a^2 (1-a)} \frac{\partial}{\partial a} -\frac{1}{8 a^2(1-a)} \right] I_{0,0}(a,0) = 0\,. \label{eq:thirdPF}
\end{align}
Being a third-order equation, it admits three solutions, which correspond to the three periods of the underlying K3 surface. We refer to them as $\varpi_0$, $\varpi_1$, and $\varpi_2$. \Cref{eq:thirdPF} has a MUM point at $a=0$ and
we assume that $\varpi_0$ is the holomorphic solution close to that point.

From the theory of K3 surfaces, it is well-known that the operator \eqref{eq:thirdPF} is a symmetric square. This means that its three solutions can be written as symmetric products of the solutions of a second-order equation,
\begin{align}
    \left[  \frac{\partial^2}{\partial a^2}+\frac{(6-9 a) }{6 a(1-a) }\frac{\partial}{\partial a} -\frac{1}{16 a(1-a) } \right] \pi_i = 0\,. \label{eq:secondPF}
\end{align}
We refer to the two solutions of this second-order Picard-Fuchs equation as $\pi_0$ and $\pi_1$.
From~\cref{eq:secondPF}, it is easy to see that they satisfy the relation
\begin{align}
    \Delta_2 = \pi_0 \pi_1' - \pi_0' \pi_1 
    = \frac{c}{a \sqrt{1-a}} \, ,
    \label{eq:Wr2}
\end{align}
and we normalize them such that $c = 1$.
With this, the three solutions of~\cref{eq:thirdPF} can be chosen as
\begin{align}
    \varpi_0 = \pi_0^2\,, \quad \varpi_1 = \pi_0 \pi_1\,, \quad 
    \varpi_2 = \frac12\pi_1^2\,. \label{eq:periodsk3}
\end{align}

For the following discussion, it is useful to list some of the relations satisfied by the (quasi-)periods of the K3 surface. 
By introducing the notation $\Delta_{ij} = \varpi_i \varpi_j' - \varpi_j \varpi_i'$ for the minors of the period matrix, it is easy to prove the following quadratic relations
\begin{align}
    \Delta_{01} &= \varpi_0 \varpi_1' - \varpi_0' \varpi_1 = \Delta_2\, \varpi_0 \, , \nonumber \\
    \Delta_{02} &= \varpi_0 \varpi_2' - \varpi_0' \varpi_2 =  \Delta_2\, \varpi_1 \, , \nonumber \\
    \Delta_{12} &= \varpi_1 \varpi_2' - \varpi_1' \varpi_2 = \Delta_2\, \varpi_2 \, .
\label{eq:quadratic}
\end{align}
Moreover, to simplify the structure of the differential equations, we will also need a quadratic relation to express the second derivative of the holomorphic period in terms of the period and its first derivative,
\begin{equation}
    -4 (2-3 a) \varpi _0 \varpi _0'+4 a(1-a) \,  \varpi _0'^2
    -8a (1-a) \,  \varpi_0 \varpi _0''+\varpi _0^2=0 \, .\label{eq:griffom2}
\end{equation}
From these we can also derive
\begin{align}
    \frac\partial{\partial a} \left( \frac{\Delta_{02}}{\Delta_{01}}\right) 
    = \frac\partial{\partial a} \left( \frac{\varpi_1}{\varpi_0}\right)
    =  \frac{ \varpi_0 \varpi_1' - \varpi_0' \varpi_1}{\varpi_0^2} 
    =  \frac{\Delta_{01}}{\varpi_0^2} 
    =  {\Delta_2}{\varpi_0} 
    = \frac{1}{ a \sqrt{1-a}\,\varpi_0}\,.
\end{align}

\subsection{Constructing a canonical basis}
The logic now exactly parallels the elliptic case.
The only novelty is that the relevant cohomology is three-dimensional. Starting from the holomorphic master we need two derivatives and therefore expect two weight drops.
This suggests the following starting basis
\begin{align}
    J_1 &= \frac{1}{\varpi_0} I_{0,0}\,, \nonumber \\
    J_2 &= \frac{1}{\epsilon}\, \frac{\varpi_0^2}{\Delta_{01}} \partial_a J_1
    = \frac{1}{\epsilon} \left( - \frac{\varpi_0'}{\Delta_{01}} I_{0,0} +  
    \frac{\varpi_0}{\Delta_{01}} I_{0,0}' \right)\,, \nonumber \\
    J_3 &= \frac{1}{\epsilon}  \, \left[\partial_a \left( \frac{\Delta_{02}}{\Delta_{01}}\right)\right]^{-1} \partial_a J_2
    = \frac{1}{\epsilon^2}
    \left[\partial_a \left( \frac{\Delta_{02}}{\Delta_{01}}\right)\right]^{-1} \partial_a \left( \frac{\varpi_0^2}{\Delta_{01}} \partial_a J_1 \right)\,.
\end{align}

By analyzing it on the three fundamental cycles $C_0,C_1,C_2$ we find
\begin{align}
    J_1 & \longrightarrow \left\{ 
    \begin{array}{clll}
         \text{1st cycle:} & \quad \text{LS}[J_1]_{C_0} =& 1   + \mathcal{O}(\epsilon)\,, & \\ \\
         \text{2nd cycle:} &  \quad\text{LS}[J_1]_{C_1} =& \frac{\varpi_1}{\varpi_0} + \mathcal{O}(\epsilon) 
         &= \tau  + \mathcal{O}(\epsilon)\,,
         \\ \\
         \text{3rd cycle:} & \quad \text{LS}[J_1]_{C_2} =& 
         \frac{\varpi_2}{\varpi_0}
         + 
         \mathcal{O}(\epsilon)
         &
         =\frac12 \tau^2
         + \mathcal{O}(\epsilon)\,,
    \end{array}
     \right. \label{eq:LS1}
\\ 
    J_2 &\longrightarrow \left\{ 
    \begin{array}{ccllcc}
         \text{1st cycle:} &  \quad\text{LS}[J_2]_{C_0} =& \mathcal{O}(1) \,,
         & & & \\ \\
         \text{2nd cycle:} & \quad \text{LS}[J_2]_{C_1} =&\frac{1}{\epsilon} \frac{\varpi_0 \varpi_1'-\varpi_0' \varpi_1 }{\Delta_{01}}  + \mathcal{O}(1) 
         &= \frac{1}{\epsilon} + \mathcal{O}(1)\,, & &
         \\ \\
         \text{3rd cycle:} & \quad\text{LS}[J_2]_{C_2} =& \frac{1}{\epsilon} \frac{ \varpi_0 \varpi_2'-\varpi_0' \varpi_2 }{\Delta_{01}}  + \mathcal{O}(1) 
          &= \frac{1}{\epsilon} \tau + \mathcal{O}(1) \,,
    \end{array}
     \right. \label{eq:LS2}
\\     
    J_3 &\longrightarrow \left\{ 
    \begin{array}{cclccc}
         \text{1st cycle:} &  \quad\text{LS}[J_3]_{C_0} =& \mathcal{O}\left( \frac{1}{\epsilon}\right) \,,
         & & & \\ \\
         \text{2nd cycle:} & \quad\text{LS}[J_3]_{C_1} =& \mathcal{O}\left( \frac{1}{\epsilon}\right) \,, 
         & & &
         \\ \\
         \text{3rd cycle:} & \quad\text{LS}[J_3]_{C_2} =& \frac{1}{\epsilon^2} 
         \left[\partial_a \left( \frac{\Delta_{02}}{\Delta_{01}}\right)\right]^{-1}
         \partial_a\left( \frac{\Delta_{02}}{\Delta_{01}} \right) &=
         \frac{1}{\epsilon^2} + \mathcal{O}\left( \frac{1}{\epsilon}\right) \,,
    \end{array}
     \right. \label{eq:LS3}
\end{align}
where in the last line we used that $\varpi_2/\varpi_0 = \tau^2/2$.
Thus $J_1$, $J_2$, and $J_3$ are normalized correctly at their first non-trivial $\epsilon$-orders, but not yet at higher orders in $\epsilon$.

The corresponding differential equations take the form
\begin{align}
    \frac\partial{\partial a} \begin{pmatrix}
        J_1 \\ J_2 \\ J_3
    \end{pmatrix}
    = \left[ \frac{1}{\epsilon} B_{-1} + B_0 + \epsilon B_1 \right]\begin{pmatrix}
        J_1 \\ J_2 \\ J_3
    \end{pmatrix}\,,
\label{eq:diffeqK3v2}
\end{align}
with
\begin{align}
    &B_{-1} = \begin{pmatrix}
        0 & 0 & 0 \\
        0 & 0 & 0 \\
        -\frac{a (2+a) \varpi _0 \varpi _0'}{4 (1-a)}
 -\frac{1}{4} a(4-a)  \left(\varpi _0'\right){}^2
 -\frac{(2+a)\varpi _0^2}{16 (1-a)} & 0 & 0
    \end{pmatrix}\,, \nonumber \\
    & B_{0} = \begin{pmatrix}
        0 & 0 & 0 \\
        0 & 0 & 0 \\
        -\frac12\left(4+a\right) \varpi _0 \varpi _0'
 -\frac{\varpi _0^2}{4}
 & -\frac{(4-a) \varpi _0'}{2\sqrt{1-a}}
 -\frac{(2+a) \varpi _0}{4 (1-a)^{3/2}} & 0
    \end{pmatrix}\,, \nonumber \\
   &  B_{1} = \begin{pmatrix}
        0 & \frac{1}{\sqrt{1-a} a \varpi _0} & 0 \\
        0 & 0 & \frac{2}{\sqrt{1-a} a \varpi _0} \\
        -\frac{\varpi _0^2}{2} & -\frac{(a+4) \varpi _0}{2 \sqrt{1-a} a} & -\frac{4-a}{(1-a) a}
    \end{pmatrix} \, .
\end{align}
Equivalently, this basis is the one obtained from the derivative basis after splitting the period matrix into semi-simple and unipotent parts, rotating by the inverse semi-simple piece, and rescaling by the appropriate powers of $\epsilon$.
Indeed, for the period matrix
\begin{align}
    W = \begin{pmatrix}
        \varpi_0 & \varpi_1 & \varpi_2 \\
        \varpi_0' & \varpi_1' & \varpi_2' \\
        \varpi_0'' & \varpi_1'' & \varpi_2''
    \end{pmatrix}
    = W_\text{ss}
    \begin{pmatrix}
        1 & \tau & \frac12\tau^2 \\
        0 & 1 & \tau \\
        0 & 0 & 1
    \end{pmatrix},
\end{align}
the unipotent part is built from the leading singularities in \cref{eq:LS1,eq:LS2,eq:LS3}.

To complete our canonical basis, we normalize the remaining leading singularities.
We start with $J_3$, which is only properly normalized at order $\epsilon^{-2}$.
By expanding each master
\begin{align}
    J_i = \sum_{j=-2}^\infty J_i^{(j)}\, \epsilon^j \,,
\end{align}
we see that the differential equations in~\cref{eq:diffeqK3v2} imply 
\begin{align}
\frac{\partial}{\partial a} J_3^{(-1)} &= \left(
 -\frac{a (2+a) \varpi _0 \varpi _0'}{4 (1-a)}
 -\frac{a(4-a) }{4}  \left(\varpi _0'\right){}^2-\frac{(2+a)
   \varpi _0^2}{16 (1-a)} \right) J_1^{(0)} 
   \nonumber \\ &
\qquad- \frac{(2+a) \varpi _0}{4 (1-a)^{3/2}} J_2^{(-1)}
- \frac{\varpi_0^2}{2} J_3^{(-2)}\,.
\end{align}
Let us focus on the first cycle $C_0$, since as we have seen, once the LS associated to that cycle is normalized, the remaining ones fall naturally into place.
Using~\cref{eq:LS1,eq:LS2} we find
\begin{align}
     \frac{\partial}{\partial a} \text{LS}[J_3^{(-1)}]_{C_0} 
   &=\frac{\partial}{\partial a} \left( -\frac{1}{6} a\left(4-a \right) \varpi _0 \varpi _0'-\frac{a\left(2+a \right) \varpi _0^2}{12 (1-a)} \right)
   + \frac{(1+a) \varpi _0^2}{8 (1-a)^2}\,,
\end{align}
which shows that by defining
\begin{align}
    K_3 = J_3 - \frac{1}{\epsilon}\, \left(  -\frac{1}{6}a \left(4-a \right) \varpi _0 \varpi _0'-\frac{a\left(2+ a\right) \varpi _0^2}{12 (1-a)} 
    + G_1(a) \right) J_1   
\end{align}
with
\begin{align}
    G_1(a) = \frac{1}{8}\, \int \mathrm d a \,\frac{(1+ a)\varpi _0^2 }{ (1- a)^2}\,, \label{eq:G1def}
\end{align}
we have
\begin{align}
    \frac{\partial}{\partial a}\text{LS}[K_3^{(-1)}]_{C_0} = 0 \quad \Rightarrow \quad
    \text{LS}[K_3^{(-1)}]_{C_0} = \text{const} \,.
\end{align}
Indeed, $K_3$ has now a constant leading singularity on the first cycle at order $1/\epsilon$, as desired.

We now have a set of masters $J_1$, $J_2$, and $K_3$ whose leading singularities are properly normalized up to order $\epsilon^{-1}$. 
Still, $J_2$ and $K_3$ do not have properly normalized leading singularities at order $\epsilon^0$ and, as a consequence, their differential equations are not yet $\epsilon$-factorized. Instead they read
\begin{align}
     \frac{\partial}{\partial a} \begin{pmatrix}
        J_1 \\ J_2 \\ K_3
    \end{pmatrix} 
    &= \big(  M_0 + \epsilon\, M_1 \big)\begin{pmatrix}
        J_1 \\ J_2 \\ K_3
    \end{pmatrix}\,,  \label{eq:diffeqK3v3}
\end{align}
where
\begin{equation}
\begin{aligned}  
&M_{0} = \begin{pmatrix}
        0 & 0 & 0 \\
        \frac{2 G_1}{\sqrt{1-a} a \varpi _0}-\frac{(4-a) \varpi _0'}{3 \sqrt{1-a}}-\frac{(2+a) \varpi _0}{6
   (1-a)^{3/2}} & 0 & 0 \\
        \frac{\left(4+a+4a^2\right) \varpi _0 \varpi _0'}{6(1-a)}+\frac{\left(5+8a-4a^2\right) \varpi _0^2}{12
   (1-a)^2}-\frac{(4-a) G_1}{(1-a) a} & -\frac{G_1}{\sqrt{1-a} a \varpi _0}-\frac{(4-a) \varpi _0'}{3
   \sqrt{1-a}}-\frac{(2+a) \varpi _0}{6 (1-a)^{3/2}} & 0
    \end{pmatrix}\,,\\
& M_{1} = \begin{pmatrix}
        0 & \frac{1}{\sqrt{1-a} a \varpi _0} & 0 \\
        0 & 0 & \frac{2}{\sqrt{1-a} a \varpi _0} \\
        -\frac{\varpi _0^2}{2} & -\frac{(4+a) \varpi _0}{2 \sqrt{1-a} a} & -\frac{4-a}{(1-a) a}
        \end{pmatrix}\,.
\end{aligned}
\end{equation}

To proceed, we need to fix the leading singularities of the second and third master at $\epsilon^0$. 
From~\cref{eq:diffeqK3v3}, we can immediately read off the differential equation satisfied by the leading singularity of the second master on the first cycle at order $\epsilon^0$,
\begin{align}
     \frac{\partial}{\partial a} \text{LS}[J_2^{(0)}]_{C_0} 
   &=\frac{\partial}{\partial a} \left( -\frac{4-a}{3 \sqrt{1-x}}\varpi_0 \right) + \frac{2 G_1}{a\sqrt{1-a}\, \varpi_0}\,,
\end{align}
where $G_1$ was defined in~\cref{eq:G1def} and, again, we wrote the right-hand side explicitly separating the part that can be written as a total derivative.
This result suggests the redefinition
\begin{align}
    K_2 = J_2 - \left(  -\frac{4-a}{3 \sqrt{1-x}}\varpi_0  + 2\, G_2(a) \right) J_1\,,
\end{align}
where
\begin{equation}
    G_2(a) = \int \mathrm d a \frac{1}{ a\sqrt{1- a}}\, \frac{G_1}{\varpi_0}\,, \label{eq:G2def}
\end{equation}
such that the leading singularity of the second master at order $\epsilon^0$ is now constant. We have
\begin{align}
     \frac{\partial}{\partial a}\text{LS}[K_2^{(0)}]_{C_0} = 0 \quad \Rightarrow \quad
    \text{LS}[K_2^{(0)}]_{C_0} = \text{const}\,.
\end{align}
After this redefinition, the differential equations take the form
   \begin{align}
    \frac{\partial}{\partial a} \begin{pmatrix}
        J_1 \\ K_2 \\ K_3
    \end{pmatrix} 
    &= \big(  D_0 + \epsilon\, D_1 \big)\begin{pmatrix}
        J_1 \\ K_2 \\ K_3
    \end{pmatrix} \,, \label{eq:diffeqK3v4}
\end{align}
where the only matrix that matters for the following discussion is $D_0$, which reads
\begin{align}
    \scalebox{1.00}{$D_0= \begin{pmatrix}
     0 & 0 & 0 \\
 0 & 0 & 0 \\
 [D_0]_{31} & [D_0]_{32} & 0
    \end{pmatrix} $}
\end{align}
with
\begin{equation}
\begin{aligned}
    [D_0]_{31} &=  \left(\frac{\left(44-13a+14a^2\right) \varpi _0'}{18(1-a)}
-\frac{(2+a) G_2}{6 (1-a)^{3/2}}\right)\varpi_0-\frac{(4-a)
   G_2 \varpi _0'}{3 \sqrt{1-a}}-\frac{G_1 G_2}{\sqrt{1-a} a \varpi _0} \nonumber \\ 
         &\qquad -\frac{2 (4-a) G_1}{3 (1-a) a}+\frac{(31+14 (2-a) a)
   \varpi _0^2}{36 (1-a)^2}\, ,
\\
[D_0]_{32} &= -\frac{G_1}{\sqrt{1-a} a \varpi _0}
-\frac{(4-a) \varpi _0'}{3 \sqrt{1-a}}-\frac{(2+a) \varpi_0}{6 (1-a)^{3/2}}\,.
\end{aligned} 
\end{equation}
From these equations, we can finally read off the differential equation satisfied by the leading singularity of the third master  on the first cycle at order $\epsilon^0$,
\begin{align}
     \frac{\partial}{\partial a} \text{LS}[K_3^{(0)}]_{C_0}  
   &= \frac{\partial}{\partial a}\left(  -\frac{(4-a) G_2 \varpi_0}{3 \sqrt{1-a}}+\frac{44-13a+14a^2}{36 (1-a)}\varpi_0^2 -\frac{1}{4} G_2^2\right) \nonumber \\
   &\qquad+ \frac{\partial}{\partial a} \left(  -\frac{4-a}{3 \sqrt{1-x}}\varpi_0  \right)
   - \frac{G_1}{a \sqrt{1-a}\, \varpi_0}\,,
\end{align}
where, once more, we explicitly reorganized all terms that could be expressed as a total derivative.
This equation suggests the final shift
\begin{align}
    L_3 = K_3 &- \left( -\frac{(4-a) G_2 \varpi_0}{3 \sqrt{1-a}}+\frac{44-13a+14a^2}{36 (1-a)}\varpi_0^2 
    -\frac{1}{4} G_2^2 \right)J_1 \nonumber \\
    &- \left( -\frac{4-a}{3 \sqrt{1-a}}\varpi_0 - G_2\right)K_2\,,
\end{align}
which guarantees
\begin{align}
    \frac{\partial}{\partial a} \text{LS}[L_3^{(0)}]_{C_0} = 0 \quad \Rightarrow \quad
    \text{LS}[L_3^{(0)}]_{C_0} = \text{const}\,.
\end{align}

At this point, all leading singularities on the first cycle $C_0$ are properly normalized to constants up to order $\epsilon^0$. 
One might wonder, what guarantees that this procedure is sufficient and that we do not need to normalize extra leading singularities at higher orders in $\epsilon$.
The crucial point, which we repeat here, is that we have started our analysis from the right integrals: We picked as first integral a candidate with at most single poles (including the contribution from the first-kind form), and as further candidates its suitably normalized derivatives. 
The fact that we differentiated twice a \emph{good integral}, guarantees that we only need to fix the resulting leading singularities to two extra $\epsilon$-orders.
This statement is easily verified by deriving the system of differential equations satisfied by the last basis we found, which is now in fully $\epsilon$-factorized form
\begin{align}
    \frac{\partial}{\partial a} \begin{pmatrix}
        J_1 \\ K_2 \\ L_3
    \end{pmatrix}
    = \epsilon\, M\, \begin{pmatrix}
        J_1 \\ K_2 \\ L_3
    \end{pmatrix} \, ,
\end{align}
where
\begin{align}
    M =  \scalebox{0.8}{$\begin{pmatrix}
\frac{\sqrt{1-a} G_2}{(1-a) a \varpi _0}-\frac{4-a}{3 (1-a) a} & \frac{1}{\sqrt{1-a} a \varpi _0} & 0 
 %==================
\\
 %==================
 \frac{\left(4+a+4a^2\right) \varpi _0}{6 (1-a)^{3/2} a}-\frac{3 G_2^2}{2 \sqrt{1-a} a \varpi _0} & 
 -\frac{2 G_2}{\sqrt{1-a} a \varpi _0}-\frac{4-a}{3 (1-a) a} & 
 \frac{2}{\sqrt{1-a} a \varpi _0} 
 %==================
 \\
 %==================
 \frac{\left(4+a+4a^2\right) G_2 \varpi _0}{6 (1-a)^{3/2} a}+\frac{\left(16-75a+21a^2-16a^3\right) \varpi _0^2}{54 (1-a)^2 a}-\frac{G_2^3}{2 \sqrt{1-a} a \varpi _0} & 
 \frac{\left(4+a+4a^2\right) \varpi _0}{12 (1-a)^{3/2}a}-\frac{3 G_2^2}{4 \sqrt{1-a} a \varpi _0} & 
   -\frac{\sqrt{1-a} G_2}{(a-1) a \varpi _0}-\frac{4-a}{3 (1-a) a}
    \end{pmatrix}$} \, .
\end{align}

The important lesson is the same as in the elliptic case, but now in a setting where no explicit iterated-integral description is available.
 Starting from a good holomorphic master and its suitably normalized derivatives, one fixes the first non-trivial leading singularities and then removes the remaining subleading ones order by order in $\epsilon$.
 Because the K3 sector is generated by two derivatives of the holomorphic period, two extra normalizations are required.
 Once these are implemented, the resulting basis is $\epsilon$-factorized.
As we stressed at various points of our construction, our procedure is 
expected to converge if one starts from a (set of) master integral(s) with at most single poles and which map to the first- and third-kind form(s) on the geometry considered, and if one picks as representative of the remaining elements of the cohomology (i.e. the second-kind forms) a suitable number of iterated derivatives of the corresponding first-kind ones.

As a final comment, we repeat once more that 
while we only worked out examples related to elliptic 
and K3 geometries, our construction only requires to properly enumerate independent cycles and differential forms, which can be generalized to arbitrary geometries in the context of mixed Hodge theory~\cite{Duhr:2025lbz}.

\section{Leading singularities arising from subtopologies}
\label{sec:subtop}

In the previous sections, we illustrated how achieving an $\epsilon$-factorized canonical basis on general geometries is equivalent to identifying integrals with unit leading singularities. 
When forms of the second kind are involved, we demonstrated that this can be achieved by normalizing the leading singularities of the corresponding integrals to higher orders in the dimensional regulator $\epsilon$.
Up to this point, however, we have, strictly speaking, worked with toy models that mimic the ``maximal cut'' of Feynman integrals, or, in other words, we have only considered the homogeneous part of the differential equations. 
This was sufficient to illustrate the main concepts.
In this section, we would like to consider more realistic examples corresponding to multi-loop Feynman integrals and show how the same analysis can be carried out in the very same way beyond the maximal cut. 
Conceptually, nothing changes, and also in these cases, new transcendental functions appear when we attempt to cast the differential equation system in $\epsilon$-form. 
Indeed, these functions can be related to extra leading singularities beyond the maximal cut. 
Working at the integrand level, we will illustrate this explicitly with one example at two loops and another at three loops, where these functions are related to third-kind forms defined on the geometry of one of their subtopologies.
In particular, in the first case we consider an integral which is polylogarithmic on its maximal cut, but couples via a subsector to an elliptic geometry. In the second, 
we consider instead a Feynman graph which is elliptic on its maximal cut, and again couples to a K3 subtopology.
We have identified these examples in the study of self-energies in QED and $\phi^3$-theory. Obvious generalizations of these two examples to higher loops and more general Calabi-Yau geometries 
also play a role in the study of self-energies, but
 do not introduce qualitatively different structures compared to those found in the examples presented here. 
We leave the explicit discussion of these additional cases for a dedicated publication.
All relevant data needed to reproduce the results obtained here 
can be found as ancillary files attached to the arXiv submission of this paper.

\subsection{The two-loop eyeball graph}
\label{sec:twoloopeb}
\begin{figure}[h!]
\centering
\begin{tikzpicture}[thick, scale=0.7]

%---------------- First diagram ----------------%

\begin{scope}

  \coordinate (L) at (-2,0);
  \coordinate (R) at (2,0);
  \coordinate (B) at (0,-2.5);

  % external legs with momentum
  \draw (-3.5,0) -- (L) node[midway, above] {$p$};
  \draw (3.5,0) -- (R);
  \draw (L) -- (R);

  % single propagator to the bottom vertex
  \draw (L) -- (B) node[pos=0.55, left=4pt] {$x$};

  % point where the cut happens
  \path (L) -- (B) coordinate[pos=0.55] (cut);

  % dashed cut
  \draw[dashed]
    ($(cut)!-0.8!90:(B)$) -- ($(cut)!0.8!90:(B)$);

  % Bubble on right edge only
  \draw (B) to[bend left=50] (R);
  \draw (B) to[bend right=50] (R);

\end{scope}

%---------------- Arrow between diagrams ----------------%

\draw[->, very thick] (4.6,-1.0) -- (6.2,-1.0);

%---------------- Second diagram ----------------%

\begin{scope}[xshift=11cm, yshift=-0.8cm]

  \coordinate (L2) at (-2,0);
  \coordinate (R2) at (2,0);

  % external legs with momentum
  \draw (-3.5,0) -- (L2) node[midway, above] {$p$};
  \draw (3.5,0) -- (R2);

  % 2-loop sunrise
  \draw (L2) to[bend left=80] (R2);
  \draw (L2) -- (R2);
  \draw (L2) to[bend right=80] (R2);

\end{scope}

\end{tikzpicture}

\caption{Left: the two-loop \emph{eyeball} graph.
Right: the graph obtained by pinching the cut propagator $x$, yielding the two-loop equal-mass \emph{sunrise} which is known to be elliptic.}

\label{fig:two loop eye-ball}
\end{figure}
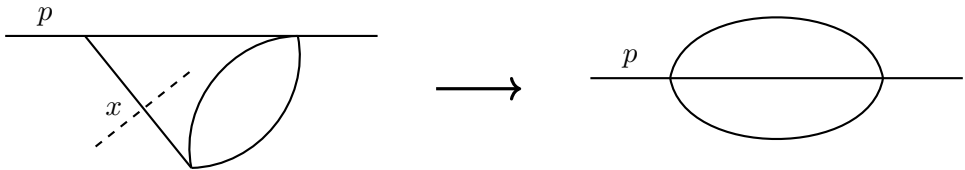

As a first example, we consider the so-called two-loop \emph{eyeball} graph~\cite{Giroux:2024yxu} shown on the left side in fig.~\ref{fig:two loop eye-ball}. We assume all propagators are massive with the same mass $m$, and we work with the dimensionless variable $z = {m^2}/{p^2}$. 
As shown on the right side of the same picture, this graph contains the two-loop sunrise as a subtopology. 
In what follows, we will analyze its leading singularities and show how an integral with unit leading singularities can be obtained. 
In doing so, we will expose a new transcendental function that represents a third-kind form on the geometry of the sunrise graph.
To keep the discussion as simple as possible, we limit ourselves to defining an integral family which is sufficient to represent the integral in a loop-by-loop Baikov representation, see~\cref{table:int family 1}. This family is not sufficient for a full IBP reduction with standard reduction tools, but it is straightforward to complete it by adding one missing irreducible scalar product.
Since we will not consider the full system of differential equations, but only focus on how to expose the non-trivial leading singularities associated with this graph, we will ignore the full integral family in what follows.
Nevertheless, we have added to this publication an ancillary file containing the definition of the complete integral family along with all details needed to construct the full canonical basis, including all subsectors.

\begin{table}[h]
\centering
\begin{tabular}{| m{2.5cm} || m{4cm} |} 
 \hline
 Denominator & Eyeball integral family  \\[5pt]
 \hline
    $x_1$ & $k_1^2 - m^2$    \\
    $x_2$ & $(k_1+k_2)^2 - m^2$  \\
    $x_3$ & $(k_2+p)^2 - m^2$   \\
    $x$   & $k_2^2 - m^2$  \\  
 \hline
\end{tabular}

\caption{The two-loop eyeball propagators used to build the loop-by-loop Baikov representation. There is an incoming off-shell momentum $p$, with $p^2 \neq 0$, and we work with the dimensionless variable $z = m^2/p^2$. The sector in fig.~\ref{fig:two loop eye-ball} corresponds to active propagators $x_1,x_2,x_3$ and $x$. We denote with $x$ the propagator that does not have support on the sunrise cut.}
\label{table:int family 1}
\end{table}

An IBP reduction for this graph exposes $6$ master integrals: The eyeball top sector contains one master integral, while the sunrise subsector contains two masters. 
The remaining three masters are grouped within simple polylogarithmic subsectors.
We will ignore all polylogarithmic subsectors, as they can be treated with standard techniques, and focus instead on the eyeball and its coupling to the sunrise graph.
In particular, by studying the integrand of the eyeball on the maximal cut, we will first deal with the homogeneous part of the differential equation, exactly as discussed in the previous sections. 
We will then focus on the non-trivial coupling with the sunrise subsector. 
To indicate the integrals in this family, we use the common notation
\begin{equation}
    I_{n_1 \cdots n_4} = \int \frac{\mathrm d^d k_1}{i \pi^{d/2}}\frac{\mathrm d^d k_2}{i \pi^{d/2}} \frac{1}{x_1^{n_1} x_2^{n_2} x_3^{n_3} x^{n_4}} \,,
\end{equation}
and we work close to two dimensions, i.e., we set $d=2-2\epsilon$.

Let us start off by considering the corner integral for the eyeball graph, $\tilde{I} = I_{1111}$.
We construct a loop-by-loop Baikov representation~\cite{Frellesvig:2024ymq} based on the propagators defined in~\cref{table:int family 1}.
After cutting the three propagators which correspond to the sunrise graph and putting $\epsilon = 0$, we obtain
\begin{align}
    \mathrm{Cut}_\text{Sun}[\tilde{I}]
        &=
    z \int \mathrm dX \, \frac1{1-X}\frac1{\sqrt{P_4^\text{Sun}(X;z)}} \, ,
\label{eq: EyeBall_subtopo_integrand}
\end{align}
where
\begin{align}
    P_4^\text{Sun}(X;z)
    &= X(4-X)Q(X,z)    \quad\text{with}\quad Q(X,Y)= 1-2(X+Y)+(X-Y)^2 \, .
\label{eq:T_subtopo_sun}
\end{align}
Notice, that we have defined $X = x+1$ in order 
to recover the standard form of the elliptic curve of the sunrise.
Additionally, we used the notation $\mathrm{Cut}_\text{Sun}$ to indicate the generalized unitarity cut of the top sector which projects onto the sunrise subtopology.

From here, it is easy to study the geometry associated with the maximal cut of the top-sector and compute its leading singularity by simply localizing all remaining propagators.
In this parametrization, this corresponds to taking the remaining residue around $X =1$,
\begin{align}
    \mathrm{LS}[\tilde{I}]_{X= 1} \sim \frac{z}{\sqrt{1-4 z }}\,,
\label{eq:eyeball_maxcut}
\end{align}
which can be normalized to one by simply rescaling the original integral as
\begin{align}
    I_1 = \frac{\sqrt{1-4 z }}{z}\,\tilde{I}.
    \label{eq: I1_eyeball}
\end{align}
Indeed, this demonstrates that the geometry underlying 
the maximal cut of this graph is the Riemann sphere and that this candidate, on the maximal cut, has unit leading singularities. 

Let us now focus  on the sunrise subtopology. 
As already discussed at length, the first crucial step is to identify an integral with at most single poles, and that contains the form of the first kind. 
A good candidate is $J_1 = I_{1110}$.\footnote{We stress that, for the sunrise, this is true beyond its maximal cut. We do not repeat the analysis here as it has already been presented in many references, including~\cite{Gorges:2023zgv, Duhr:2025lbz}} 
From our discussion above, we also know that we can pick the second starting integral as the derivative of the first, i.e., $J_2 = \partial_z J_1$. 
With this choice, let us focus on the block $\vec{J} = ( J_1, J_2, \tilde{I})$ satisfying the differential equations
\begin{align}
\frac\partial{\partial z} \vec{J} = M \cdot \vec{J}
\end{align}
with connection matrix given by
\begin{align}
M =
\begin{pmatrix}
0 & 1 & 0 \\[6pt]
-\frac{(1 - 3z) +  (3 - 5z)\epsilon + 2(1 + z)\epsilon^2}{z^2(1-z)(1-9z)}
&
\frac{(1 - 9 z^2) +  (3 - 10 z - 9 z^2)\epsilon}{z(1-z)(1-9z)}
&
0 \\[10pt]
\frac{2+\epsilon}{3(4z-1)}
&
\frac{5z}{3(1-4z)}
&
\frac{1-2z+\epsilon}{z\,(1-4z)}
\end{pmatrix},
\end{align}
which is clearly not $\epsilon$-factorized.

The rotation of the  $2\times2$ block of the sunrise graph to a canonical $\epsilon$-factorized form is well understood and can be achieved following the discussion in~\cref{sec:ell} and appendix~\ref{app:sunrise}.
Combining~\cref{eq: I1_eyeball} with the rotation to put the sunrise $2 \times 2$ block in canonical form, we get the change of basis $\vec{J}^{\,\prime} = R\cdot\vec{J}$, where
\begin{align}
R =
\begin{pmatrix}
\frac{\epsilon}{\varpi_0} & 0 & 0 \\[10pt]
\frac{(3-10z-9z^2)\,\epsilon\,\varpi_0}{4z^2}
+\frac{(1 - z) (1 - 9 z)\,\varpi_0'}{2z}
&
-\frac{(1 - z) (1 - 9 z)\,\varpi_0}{2z}
&
0 \\[10pt]
0 & 0 & \epsilon \frac{\sqrt{1-4 z}}{z}
\end{pmatrix} \, .
\label{eq: rotation 1}
\end{align}
Here, $\varpi_0$ is the holomorphic period of the sunset elliptic curve defined in~\cref{eq:T_subtopo_sun}.
After this rotation, the new differential equation for the rotated basis $\vec{J}^{\,\prime}$
reads
\begin{equation}
    \frac\partial{\partial z}\vec{J}^{\,\prime} = (A+\epsilon B) \vec{J}^{\,\prime}\,, 
\end{equation}
where
\begin{align}
A =
\begin{pmatrix}
0 & 0 & 0 \\
 0 & 0 & 0 \\
 \frac{2 \varpi_0-5 z \varpi_0'}{3 z \sqrt{1-4 z}} & 0 & 0
\end{pmatrix}\,, \quad
B =
\begin{pmatrix}
 \frac{3-10 z-9 z^2}{2 z(1-z)  (1-9 z)} & -\frac{2 z}{ (1-z) (1-9
   z)\varpi_0^2} & 0 \\
 -\frac{ (1+3 z)^4\varpi_0^2}{8z^3 (1-z)  (1-9 z)} & \frac{3-10 z-9 z^2}{2
   z(1-z)  (1-9 z)} & 0 \\
 -\frac{\varpi_0 \left(13-30 z-63 z^2\right)}{6z (1-z)   (1-9 z)\sqrt{1-4 z}} &
   \frac{10 z}{3  (1-z)  (1-9 z)\sqrt{1-4 z}\varpi_0} & 
   \frac{1}{z (1-4 z)} 
\end{pmatrix}\,.
\label{eq: B_z}
\end{align}
A leftover term remains at order $\epsilon^0$ in the coupling of the eyeball with the sunrise. This can be rewritten as
\begin{align}
     \frac{1}{3 z\sqrt{1-4 z}}\left(2 \varpi_0 - 5 z \,\varpi'_0 \right)  &= \frac{2 (1+z)\varpi_0}{3 z (1 -4 z) \sqrt{1-4 z}} - \frac{\partial}{\partial z} \left(  \frac{5\varpi_0}{3\sqrt{1-4 z}}\right)\nonumber \\
     &= -\frac{\partial}{\partial z} \left(G(z) +  \frac{5\varpi_0}{3\sqrt{1-4 z}} \right) \, .
\label{eq: ep_0 eye ball sub topo}
\end{align}
where we have defined  $G(z)$ as the function satisfying the differential equation
\begin{align}
    G'(z) = \frac{2 (1+z)\varpi_0}{3 z (1-4 z)^{3/2}}\,.
    \label{eq: G_function_2loop_eyeball}
\end{align}
This suggests to perform a second (\emph{clean-up}) 
rotation by subtracting a contribution proportional to the sunrise,
\begin{align}
    I = I_1 -  \left(G(z) +  \frac{5\varpi_0}{3\sqrt{1-4 z}} \right) \frac{J_1}{\varpi_0}.
    \label{eq: eyeball_canonical1}
\end{align}
One can easily verify that the differential equation satisfied by $I$ is now fully $\epsilon$-factorized with the newly defined $G$-function entering in the coupling to the sunrise subtopology.

\subsubsection*{The origin of G as a leading singularity of $I_1$}

Our goal in this section is to understand the rotation in~\cref{eq: eyeball_canonical1} from the perspective of leading singularities. 
In particular, we want to show that the term in brackets removes an additional leading singularity, allowing us to define an integral with unit leading singularities.
Let us go back to~\cref{eq: EyeBall_subtopo_integrand}.
While the LS corresponding to the integration contour around the residue at $X=1$ is properly normalized to $1$, the square root in~\cref{eq:T_subtopo_sun} introduces additional independent cycles and therefore additional LS according to~\cref{sec:idea}.
In particular, the integrand in eq.~\eqref{eq: EyeBall_subtopo_integrand} corresponds to a form of the third kind defined on the elliptic geometry of the maximal cut of the two-loop sunrise. 
Therefore, according to our prescription, in addition to the contour encircling the pole, we also need to consider the LS associated with the $A$-cycle of the sunrise elliptic curve. 

Let us define
\begin{align}
    \tilde{G} =  z\oint_A \mathrm dX\;
    \frac1{1-X}\frac1{\sqrt{P_4^\text{Sun}(X;z)}} \, .
\end{align}
To show that~\cref{eq: eyeball_canonical1} exactly removes the leading singularity given by $\tilde G$, we will derive the differential equation satisfied by $\tilde{G}$  and show that, after accounting for the overall normalization in~\cref{eq: I1_eyeball}, this differential equation indeed corresponds to~\cref{eq: G_function_2loop_eyeball}.
The simplest way to derive this differential equation is to reconstruct it from a series expansion for $\tilde{G}$ at a singular point, which we pick to be $z=0$ for definiteness.
In particular, it is easiest to use the holomorphic solution, i.e., the solution around the $A$-cycle, which we can obtain  by expanding  the integrand in eq.~\eqref{eq: EyeBall_subtopo_integrand} as a power series in $z$. 
This corresponds to the computation of the so-called \emph{torus period}, or, in the language of expansion by regions, on the \emph{hard branch} of $\tilde{G}$ as $z\sim0$ (see also section 7 in~\cite{Duhr:2025lbz}, where this procedure was also applied). 
Up to an irrelevant numerical normalization, we then easily find
\begin{align}
\tilde{G}
& \sim
z \int \mathrm dX\, 
\frac{1+(1+X)z+(1+4X+X^2)z^2+(1+9X+9X^2+X^3)z^3+\mathcal O(z^4)}{(1-X)\sqrt{X(4-X)}} \, .
\label{eq:EyeBall_subtopo_expanded}
\end{align}
The integrand in~\cref{eq:EyeBall_subtopo_expanded} has two single poles, one at $X=1$ which corresponds to the maximal cut, and another at $X=\infty$, which corresponds to the $A$-cycle that we want to compute. Calculating the residue at infinity yields a series representation for $\tilde{G}$, whose first orders read
\begin{align}
 \tilde{G} \sim z\left(z + 7 z^2 + 45 z^3 + \mathcal{O}({z^4})
 \right) \, .
 \label{eq: tildeG_def_eyeBall}
\end{align}
From this series representation, we can derive the inhomogeneous differential equation
\begin{align}
    \left(1-2 z\right) \tilde{G}(z) - (1-4 z)z\, \partial_z \tilde{G}(z) = -\frac{2}{3}\varpi_0 + \frac{5}{3} z\, \varpi'_0 \, .
\label{eq: second Order operator for Gtilde sun}
\end{align}

To draw the connection between $\tilde{G}$ and the $G$ function introduced in eq.~\eqref{eq: eyeball_canonical1}, we need to take into account that in deriving $\tilde{G}$, we used $\tilde{I}$ without normalizing it by the maximal cut leading singularity. 
By including this normalization, and also taking into account the shift performed in the second line of eq.~\eqref{eq: ep_0 eye ball sub topo}, we get the identity
\begin{align}
     \frac{\sqrt{1-4 z }}{z} \tilde{G}(z) =  G(z) + \frac{5}{3\sqrt{1-4 z}}\varpi_0(z) \, ,
     \label{eq:G function from integrand analysis}
\end{align}
where $G(z)$ satisfies the same differential equation as in~\cref{eq: G_function_2loop_eyeball}.
It is important to note that the right-hand side of eq.~\eqref{eq: G_function_2loop_eyeball} is not a total derivative. 
Therefore, the function $G(z)$ represents a new and independent transcendental function that cannot be expressed in terms of the periods of the elliptic curve itself and has to be included in the rotation to a canonical basis.

In summary, the new $G$-function required to put the differential equations in $\epsilon$-factorized form can be interpreted as a new leading singularity of the eyeball integral associated with the geometry of the sunrise graph.
With~\cref{eq: eyeball_canonical1}, we effectively subtract this non-normalized LS, defining an integral with unit leading singularities. 
Explicitly, by specializing the new integral on its residue 
or the $A$-cycle of the sunrise graph, we find
    \begin{align}
       I
       \longrightarrow
       \left\{  \begin{array}{ccl}
        \text{residue}\text{:} & \text{LS}\left[I\right]_{X=1} &= 1 +\mathcal{O}(\epsilon)\, , \\
      A\text{-cycle:}& \text{LS}\left[I \right]_{A} &= \mathcal{O}(\epsilon) \, . 
        \end{array}\right.
    \end{align}

%====================================================
%====================================================
%====================================================
\subsection{A three-loop double eyeball graph}

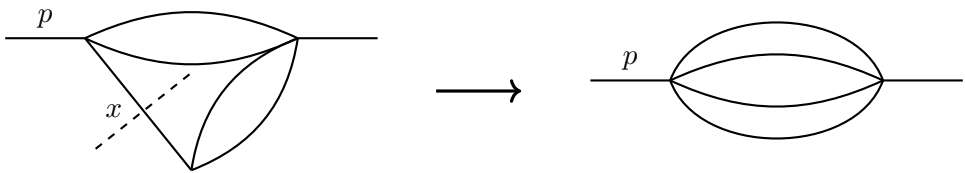
\begin{figure}[h]
\centering
\begin{tikzpicture}[thick, scale=0.7]

%---------------- First diagram ----------------%

\begin{scope}

  \coordinate (L) at (-2,0);
  \coordinate (R) at (2,0);
  \coordinate (B) at (0,-2.5);

  % external legs with momentum
  \draw (-3.5,0) -- (L) node[midway, above] {$p$};
  \draw (3.5,0) -- (R);

  \draw (L) -- (B) node[pos=0.55, left=4pt] {$x$};

  % point where the cut happens
  \path (L) -- (B) coordinate[pos=0.55] (cut);

  % dashed cut
  \draw[dashed]
  ($(cut)!-0.8!90:(B)$) -- ($(cut)!0.8!90:(B)$);

  % Bubble on right edge
  \draw (B) to[bend left=30] (R);
  \draw (B) to[bend right=30] (R);

  % Bubble on top edge
  \draw (L) to[bend left=25] (R);
  \draw (L) to[bend right=25] (R);

  \node[anchor=east] at ([xshift=-8pt]current bounding box.west) {$ $};

\end{scope}

%---------------- Arrow between diagrams ----------------%

\draw[->, very thick] (4.6,-1.0) -- (6.2,-1.0);

%---------------- Second diagram ----------------%

\begin{scope}[xshift=11cm, yshift=-0.8cm]

  \coordinate (L2) at (-2,0);
  \coordinate (R2) at (2,0);

  % external legs with momentum
  \draw (-3.5,0) -- (L2) node[midway, above] {$p$};
  \draw (3.5,0) -- (R2);

  % 3-loop banana
  \draw (L2) to[bend left=70] (R2);
  \draw (L2) to[bend left=25] (R2);
  \draw (L2) to[bend right=25] (R2);
  \draw (L2) to[bend right=70] (R2);

  \node[anchor=west] at ([xshift=8pt]current bounding box.east) {};
\end{scope}

\end{tikzpicture}

\caption{
Left: the three-loop \emph{double eyeball} graph. Right: the graph obtained by pinching the cut propagator $x$, yielding the three-loop equal-mass \emph{banana} which is related to a K3 surface.
}
\label{fig:bubble_times_bubble}
\end{figure}

We now turn to a three-loop generalization of the graph considered before, which we will refer to as \emph{double eyeball graph}, see~\cref{fig:bubble_times_bubble}. 
As before, in~\cref{table:int family 2} we limit ourselves to providing the integral family that we will use later on to derive a loop-by-loop representation, and, as in the previous section, we indicate integrals in this family as
\begin{equation}
    I_{n_1 \cdots n_6} = \int \prod_{j=1}^3 \frac{\mathrm d^dk_j}{i \pi^{d/2}} 
\frac{y^{-n_6}}{x_1^{n_1}x_2^{n_2} x_3^{n_3} x_4^{n_4} x^{n_5} }\,.
\end{equation}
Notice that we used different symbols for the last two propagators, $x,y$, as they will play a special role in the following.

\begin{table}[h]
\centering
\begin{tabular}{| m{2.5cm} || m{5.2cm} |} 
 \hline
 Denominator & Double eyeball integral family  \\[5pt]  \hline
$x_1$ & $k_1^2 - m^2$               \\
$x_2$ & $(k_1+k_2)^2 - m^2$       \\
$x_3$ & $(k_2+k_3)^2 - m^2$         \\
$x_4$ & $(k_3-p)^2 - m^2$               \\
$x$ & $k_2^2 - m^2$                  \\
$y$ & $k_3^2$                   \\ \hline
\end{tabular}
\caption{The three-loop double eyeball integral family propagators used to build the loop-by-loop Baikov representation. There is an incoming off-shell momentum $p$, with $p^2 \neq 0$, and we use again the dimensionless variable $z = m^2/p^2$. The sector in fig.~\ref{fig:bubble_times_bubble} corresponds to the active propagators $x_1,x_2,x_3,x_4,x$, while $y$ is an auxiliary irreducible scalar product.  We denote with $x$ the propagator which does not have support on the banana cut.}
\label{table:int family 2}
\end{table}

By running a standard IBP reduction code,\footnote{For this case we used Reduze2~\cite{Studerus:2009ye,vonManteuffel:2012je}.} it is easy to prove that this problem has $10$ master integrals. The top sector and the three-loop banana subtopology have $3$ masters each, for a total of 6 masters. 
The remaining $4$ master integrals are simple polylogarithmic subsectors, and we will ignore them in the following.
On the other hand, we remind the reader that the banana graph is a well-known example of a topology with underlying K3 geometry. 
In addition, as we will see, the top sector is related to an elliptic curve, and the goal of this section will be to illustrate how to find integrals with unit leading singularities accounting for the non-trivial interplay of these two geometries.

We start by constructing a loop-by-loop Baikov representation for the full graph,
using the variables in~\cref{table:int family 2}. Since we ignore the polylogarithmic subtopologies in the subsequent discussion, we immediately specialize to the banana cut at $\epsilon = 0$, which we indicate by $\mathrm{Cut}_\text{K3}$. This still gives us access to the two interesting sectors, namely the top sector and the banana subtopology.
Explicitly, we find
\begin{align}
\mathrm{Cut}_\text{K3}[I_{11111n}]
&=
z \int \mathrm dX_1\mathrm dX_2 \;
\frac1{1-X_1}\frac{X_2^{-n} }{\sqrt{P_{4,4}^\text{K3}(X_1,X_2;z)}}\,,
\label{eq:T_subtopo_integrand}
\end{align}
where
\begin{align}
P_{4,4}^\text{K3}(X_1,X_2;z)  
&=
X_1(4-X_1)Q(X_1,X_2)Q(X_2z,z) \, .
\label{eq:T_subtopo_K3}
\end{align}
Here we have again shifted the variable $X_1=x+1$, renamed $X_2=y$, and introduced the quadratic polynomial $Q(X,Y)$ defined in~\cref{eq:T_subtopo_sun}. The equation $Y^2 = P_{4,4}^\text{K3}(X_1,X_2;z)$ can be shown to define the same K3 surface
appearing in the three-loop equal-mass banana graph, and, in this section, we indicate its three periods as $\pi_0$, $\pi_1$ and $\pi_2$.

In~\cref{eq:T_subtopo_integrand}, we consider a whole family of integrals $I_{11111n}$, allowing for general powers of the irreducible scalar product $X_2 = k_3^2$.
We can project them on the maximal cut by taking the additional residue at $X_1=1$, i.e.,
\begin{align}
\mathrm{Cut}_\text{Max}[I_{11111n}]
&=
z\, \int \mathrm d X_2\;
\frac{X_2^{-n}}{\sqrt{P^\text{Sun}_4(X_2;z)}} \, ,
\label{eq:T_maxcut_integrand}
\end{align}
where $P^\text{Sun}_4(X_2;z)$ defined in~\eqref{eq:T_subtopo_sun} is the quartic polynomial defining the elliptic curve of the equal-mass sunrise graph.
We indicate the two elliptic periods as $\varpi_0$ and $\varpi_1$.

Let us now start by selecting a basis for the top sector.
Since IBP identities uncovered 3 master integrals, we expect, besides integrals representing the forms of the first and second kind, also a third master integral related to a third-kind form.
Indeed, in two dimensions, the corresponding basis is easily identified by considering~\cref{eq:T_maxcut_integrand}. One finds
\begin{align}
    \tilde{I}_1 &= I_{111110}\, , 
    \qquad 
    \tilde{I}_2 = \partial_z \tilde{I}_1\,, \qquad 
    \tilde{I}_3 = I_{11111-1}
    \label{eq: starting_basis}\,,
\end{align}
where the third integral has a single pole at infinity and provides a third-kind form. 
From this basis, we know how to obtain integrals with unit leading singularities on the maximal cut as described in the previous sections and we do not repeat the derivation here.\footnote{An alternative derivation of a canonical basis for the sunrise graph based on integrand analysis in terms of eMPLs is presented in appendix~\ref{app:sunrise}.}

Let us instead consider the non-trivial coupling of this elliptic top-sector to the three-loop banana graph, and study the new leading singularities associated with it.
As already discussed, on the maximal cut of the three-loop banana, one finds three master integrals, and the underlying geometry is a K3 surface. 
The construction of a canonical basis for the three-loop banana graph is well understood and, in the language of this paper, it follows exactly what was discussed already in~\cref{sec:K3ex}.
In particular, we start from three integrals corresponding to the first-kind form, together with its first and second derivatives
\begin{equation}
	\tilde J_1 = I_{111100}\,, \qquad \tilde J_2 = \partial_z \tilde J_1\,, \qquad \tilde J_3 = \partial_z^2 \tilde J_1 \, .
\end{equation} 
To obtain a canonical basis for them, we rescale this basis by appropriate powers of $\epsilon$ to account for the weight drops, rotate away the semi-simple part of the period matrix and perform a clean-up step as described in the previous sections (see also~\cite{Pogel:2022yat,Duhr:2025lbz}).

Once this is done, we still find non-$\epsilon$-factorized terms in the couplings among the three masters in the top topology and the ones in the banana subtopology. 
By working directly on the differential equations, these couplings can be easily removed by a further clean-up step~\cite{Gorges:2023zgv, Duhr:2025lbz}, to find a fully $\epsilon$-factorized system of differential equations. In the ancillary file to this publication, we give all ingredients to build a fully canonical differential equation, also including all other subsector contributions.

Here we focus on two of the three integrals in~\cref{eq: starting_basis}, $\tilde I_1$ and $\tilde I_3$, from which we can build the following canonical integrals 
\begin{equation}
\begin{aligned}
    I_1 &=  \frac{\tilde{I}_1}{\varpi_0}- 
    \left( \frac{G_2}{\pi_0} - \frac{5z(1-2z)}{2(1-z)(1-9z)\varpi_0} \right) \tilde J_1  \, , \\
    I_3 &= \tilde I_3 - \frac{1+3z}{3z} \tilde I_1 + \left( \frac{G_3}{\pi_0} - \frac5{6(1-z)} \right) \tilde J_1 \, .
\label{eq:T_canonical_third_firstkind}
\end{aligned}
\end{equation}
The $G_i$ are new independent transcendental functions defined by the following set of first-order differential equations
\begin{equation}
\begin{aligned}
G_1' &= \left(
    -\frac{3 - 5 z + 12 z^2}{4z^2(1 - z)(1 - 9 z)}\,\varpi_0
    +\frac{2 - 5 z + 23 z^2}{2z(1 - z)(1 - 9 z)}\,\varpi_0' \right) \pi_0 \, , \\
G_2' &=
    \frac{2 z}{(1 - z)(1 - 9 z)}\frac{G_1}{\varpi_0^2}
    -\frac{2 - 5 z + 23 z^2}{(1 - z)^2(-1 + 9 z)^2}\frac{\pi_0}{\varpi_0} \, ,\\
G_3' &=
    \frac{1 + 4 z}{6z(1 - z)^2}\,\pi_0 \, ,
\label{eq: G functions}
\end{aligned}
\end{equation}
where $G_1$ only enters into the master integrals indirectly through the definition of $G_2$.
We notice that the functions $G_i$ are generalized third-kind periods of the K3 surface.
The reason to focus on these integrals 
is that they have no weight drops and, therefore, 
it is possible to determine canonical candidates by integrand analysis exactly at $\epsilon=0$. 

On the other hand, due to the weight drop involved in its definition, the corresponding analysis for $I_2$ is more involved, and we will not reproduce it here for brevity. 
Nevertheless, we stress that it can equally well be obtained by normalizing the leading singularities of the second integral to order $\epsilon^0$, in the same spirit of our discussion in~\cref{subsec:EllipticLeadSingDEQs,subsec:Elliptic Int Analysis}, and it requires the introduction of extra transcendental $G$-functions in addition to the three introduced above.
In this case, it is indeed more convenient to perform the relevant analysis at higher orders in $\epsilon$ directly on the differential equation. As explained before, this approach is equivalent but more immediate, since differential equations already know about the full cohomology and no tailored integration by parts on the integrand are required. 

We then concentrate  on how to derive the two candidates in~\cref{eq:T_canonical_third_firstkind}
by defining combinations of integrals with unit leading singularities.
As we can easily see, the first integral only requires a correction proportional to $\tilde J_1$, whereas $I_3$ additionally also gets a shift proportional to $\tilde I_1$.
We will consider the candidates associated with $I_1$ and $I_3$ separately and show how the corresponding functions $G_2$ and $G_3$ arise from the requirement of canceling different non-constant leading singularities.

%=====================================
%=====================================
%=====================================
\subsubsection*{The origin of $G_2$ as a leading singularity of $I_1$}

We start by considering $\tilde{I}_1$ and analyzing its leading singularities.
The integrand in eq.~\eqref{eq:T_subtopo_integrand} matches that of the three-loop banana integrand on the maximal cut, with the addition of an extra simple pole at $X_1=1$. 
By taking the residue around this pole, we project the integrand on the maximal cut, where integrating a first-kind form over the two cycles generates the periods of the corresponding elliptic curve. 
This amounts to normalizing the first integral by the period of the elliptic curve, $\varpi_0$.
This is, however, not the only LS and we can isolate a second one. 
In fact, the integrand under study has non-zero support also on
the holomorphic cycle of the K3 surface $C_0$ defined via
\begin{align}
   \mathrm{LS}[J_1]_{C_0} = \oint_{C_0} J_1 = \pi_0 \, .
\end{align}
To compute this LS, we proceed similarly as for the eyeball example.
First, we compute the (holomorphic) series solution 
for $\mathrm{LS}[\tilde I_1]_{C_0}$ around the holomorphic cycle $C_0$,
\begin{equation}
\mathrm{LS}[\tilde I_1]_{C_0}
 \sim 
    z \int {\textstyle \mathrm dX_1\mathrm dX_2\;
    \frac{1+ (1+X_2)z+(1+4X_2+X_2^2)z^2 + (1+9X_2+9X_2^2+X_2^3)z^3 + \mathcal{O}(z^4)}{(1-X_1)\sqrt{X_1(4-X_1)}\sqrt{Q(X_1,X_2)}}    }\, .
\label{eq:T_subtopo_expanded}
\end{equation}
The cycle $C_0$ is defined such that we pick up the residue at infinity first in  $X_2$ and then in $X_1$ of~\eqref{eq:T_subtopo_expanded}. In this way, we obtain
\begin{align}
\tilde{G}(z) =
\mathrm{LS}[\tilde{I}_1]_{C_0} \sim
    z\left(z + 11 z^2 + 117z^3 +\mathcal{O}(z^4)\right)  \, .
\label{eq:definition Gtilde}
\end{align}
As for the previous example, we can derive an inhomogeneous differential equation for the function $\tilde G$ from its series expansion
\begin{equation}
    \mathcal L_\text{Sun}^{(2)} \tilde G = \frac z2\, \pi_0 + \frac{z^2(1-10z)}2 \, \pi_0' + \frac{5z^3(1-2z)}2 \, \pi_0'' \, ,
\label{eq:inhomG}
\end{equation}
where $\mathcal L_\text{Sun}^{(2)}$ is the second-order Picard-Fuchs operator of the sunrise graph given by
\begin{equation}
    \mathcal L_\text{Sun}^{(2)} = (1-z)(1-9z)\theta^2-2(1-5z)\theta + (1-3z) \quad\text{with}\quad \theta = z\, \partial_z \, .
\end{equation}

To show the connection between $\tilde G$ and the rotation defined in~\eqref{eq:T_canonical_third_firstkind}, we notice that using the differential equations given in~\cref{eq: G functions}, one can also derive a second-order equation for $G_2$. 
Taking into account that we have to normalize $\tilde I_1$ by the holomorphic elliptic period $\varpi_0$, one derives the same differential equation as stated in~\cref{eq:inhomG}.

Indeed, after appropriate subtraction of this new LS as in~\cref{eq:T_canonical_third_firstkind}, $I_1$ has unit leading singularities on the two holomorphic cycles related the K3 surface of the banana and the sunrise elliptic curve
\begin{align}
I_1
\longrightarrow
\left\{
\begin{aligned}
A\text{-cycle}  &:\; \mathrm{LS}\left[I_1\right]_A= 1 + \mathcal{O}(\epsilon)\, , \\
C_0\text{-cycle} &:\;\mathrm{LS}\left[I_1\right]_{C_0} = \mathcal{O}(\epsilon)\, . \\
\end{aligned}
\right.
\end{align}

%=====================================
%=====================================
%=====================================
\subsubsection*{The origin of $G_3$ as a leading singularity of $I_3$}

We now consider $\tilde{I}_3$, and demonstrate that it has a LS associated with $G_3$.
Again, we proceed as prescribed in eq.~\eqref{eq:LStksub}.
On the maximal cut, for $\epsilon = 0$, $\tilde{I}_3$ has a pole at infinity in $X_2$ which leads to a LS that is already normalized to a constant. 
Nonetheless, additional leading singularities are present due to the non-trivial cycles associated with the elliptic curve and the K3 surface. 

Let us start from the elliptic curve, which we see already on the maximal cut of the integral.
We can derive as before a series expansion for the LS defined by integrating over the $A$-cycle of the elliptic curve.
Interestingly, we find
\begin{equation}
\begin{aligned}
\mathrm{LS}[\tilde{I}_3]_A &=
1 + \frac{2}{3}z + \frac{8}{27}z^2  + \mathcal{O}(z^3) \\
&=\frac{1+3z}{3z}\, \varpi_0 =\frac{1+3z}{3z}\,\, \mathrm{LS}[\tilde{I}_1]_A \, ,
\label{eq:I3 second leading sing max cut}
\end{aligned}
\end{equation}
which shows that this LS does not give rise to a new transcendental function.
This identity justifies the first part of the definition of $I_3$ given in~\cref{eq:T_canonical_third_firstkind}.
In other words, we have
\begin{align}
    \mathrm{LS}\left[\tilde{I}_3 -  \frac{1+3z}{3z}\tilde{I}_1\right]_A= \mathcal{O}(\epsilon).
\label{eq:I3 second leading sing max cut}
\end{align}

As a second possibility, 
we need to study the LS associated with the holomorphic cycle $C_0$ of the banana graph.
To this aim, we have to compute the LS of the integral in~\cref{eq:T_subtopo_integrand} along $C_0$ for $n=-1$.
The procedure to calculate the series expansion in $z$ of this LS is the same as in~\cref{eq:T_subtopo_expanded}. We obtain
\begin{align}
    \mathrm{LS}\left[\tilde{I}_3  - \frac{1+3 z}{z}\tilde{I}_1\right]_{C_0},
    = z\left( z + 5 z^2 + 36 z^3  \mathcal{O}(z^4)\right)  = \hat G(z) \,, 
\label{eq:I3_G}
\end{align}
which satisfies
\begin{align}
    \frac\partial{\partial z} \hat G = -\frac{1}{6 (1-z)}\,\pi_0 + \frac{5z}{6(1-z)}\, \pi_0' \, .
\end{align}
This is not yet the same $G_3$ as in eq.~\eqref{eq: G functions} as we have to subtract a total derivative from it. We obtain
\begin{align}
    G_3 = \hat G - \frac{5z}{6(1-z)}\,  \pi_0 \, ,
\end{align}
which now satisfies the same differential equation as in~\eqref{eq: G functions}.
Indeed, with these subtractions we compute the LS of $I_3$ on the holomorphic cycles and the additional pole at $X_2=\infty$
\begin{align}
I_3 \longrightarrow
\left\{
\begin{alignedat}{2}
\text{$X_2=\infty$:}\quad & \mathrm{LS}[I_3]_{X_2=\infty} &={}& 1 + \mathcal{O}(\epsilon) \, , \\
\text{$A$-cycle:}\quad  & \mathrm{LS}[I_3]_{A}        &={}& \mathcal{O}(\epsilon) \, , \\
\text{$C_0$-cycle:}\quad& \mathrm{LS}[I_3]_{C_0}      &={}& \mathcal{O}(\epsilon) \, .
\end{alignedat}
\right.
\label{eq:I3_cycles_3}
\end{align}
As expected, the contributions form the $A$- and $C_0$-cycles have been completely removed, and the new LS is normalized to 1, which makes it an (independent) candidate with unit leading singularities.

% !TEX encoding = UTF-8 Unicode
% !TEX root = main.tex
\section{Conclusions}
\label{sec:con}

In this paper, we have elaborated on the concept of leading singularities beyond polylogarithms and proposed a definition of integrals with unit leading singularities, which also satisfy $\epsilon$-factorized differential equations. 
The result of our investigation is that, when considering Feynman integrals defined on geometries whose cohomology is spanned by differential forms with higher poles, the construction of a canonical basis that satisfies $\epsilon$-factorized differential equations requires analyzing their leading singularities to higher orders in $\epsilon$. 
A pole of order $n$ produces a weight drop of order $n-1$ and thus, $n-1$ additional orders in $\epsilon$ have to be considered.
This analysis is completely general and allows us to elevate our conceptual understanding of integrand analysis for Feynman integrals beyond polylogarithms to the same level as for purely polylogarithmic problems. 

We have explicitly demonstrated our argument, starting with toy models based on elliptic and K3 geometries, and then moving to two- and three-loop realistic Feynman integrals.
In these latter cases, we have analyzed the leading singularities emerging from the interplay of non-trivial geometries on the maximal cut and on the subtopologies.
Additionally, we have provided a practical way to determine the new transcendental functions that are required to put the differential equations into canonical form and interpreted them in terms of leading singularities. 
Our analysis was based on the evaluation of the integrals defining these new leading singularities.
We have done this by computing the corresponding torus period (hard branch) as a series expansion, which we then used to infer the differential equations satisfied by these functions. 
This method is completely general and only requires Taylor expanding the integrands and evaluating their residues.

We believe that our findings are conceptually important, as they provide a sought-after interpretation in terms of integrand analysis and leading singularities of various steps of the newly introduced procedures to find $\epsilon$-factorized differential equations.
In this sense, they should constitute an important piece of the puzzle to reach a consensus on a general definition of canonical integrals beyond the polylogarithmic case.
Moreover, the methods illustrated here are also of practical utility, as they provide an alternative tool to find rotations to $\epsilon$-factorized bases explicitly.

\acknowledgments
We thank C. Duhr, A. Klemm, S. Maggio, A. von Manteuffel, B. Sauer, Y. S\"ohnle for useful discussion.
This work was partly funded by the European Union through the European Research Council under the grant agreements 949279 (ERC Starting Grant HighPHun (LT, CN), 101167287 (ERC Synergy Grant  MaScAmp (CN)), by UKRI Frontier Research Grant, underwriting the ERC Consolidator Grant precSM (UKRI946 (CM)) and by the Deutsche
Forschungsgemeinschaft (DFG, German Research Foundation) through the
Excellence Cluster ORIGINS EXC-2094-390783311 (CM, FJW) and the research unit FOR 5582
– Projektnummer 508889767 (FF).
Views and opinions expressed are however those of the author(s) only and do not necessarily reflect those of the European Union or the European Research Council. Neither the European Union nor the granting authority can be held responsible for them.
CN and LT would like to thank the Erwin Schrödinger International Institute for Mathematics and Physics (ESI), University of Vienna (Austria), for the opportunity to participate in the Thematic Programme ``Amplitudes and Algebraic Geometry'' in 2026 where part of this work has been accomplished and for the support given. LT thanks the Kavli Institute for Theoretical Physics for support during the initial stages of this project.

\appendix

\section{Integrand analysis for the sunrise graph}
\label{app:sunrise}

In \cref{subsec:Elliptic Int Analysis}, we determined a canonical basis for an elliptic cubic toy model from an integrand analysis, using pure elliptic multiple polylogarithms. 
Here, we apply the same procedure to the arguably most famous elliptic Feynman integral, associated with the equal-mass sunrise graph. 
We will restrict ourselves to its maximal cut for simplicity, but an extension beyond the maximal cut is straightforward. 

For the cubic model, pure kernels were defined in the literature only on the torus (i.e., in the variable typically called $z$). 
Therefore, in~\cref{subsec:Elliptic Int Analysis}, we performed a change of variable to $z$ in order to match the pure eMPLs. 
However, since we encounter a quartic polynomial for the sunrise, we find it more convenient here to work in the variable $x$ and use the pure eMPL kernels $\Psi_i(c,x,\vec a)$ defined in~\cite{Broedel:2018qkq}.

Up to an irrelevant numerical normalization, the maximal cut of the sunrise integral in $d=2-2\epsilon$ dimensions reads
\begin{align}
    I_n^C(s,\epsilon) = \int_C \dd x\,\frac{x^n\, u^{-\epsilon }}{y}
\end{align}
with 
\begin{align}
    y = \sqrt{x(x-4)\left((1-x)^2)-2 s (x+1)+s^2\right)}\,, \quad u = \frac{(x-4) \left((1-x)^2-2 s (x+1)+s^2\right)}{s}\,.
\end{align}
We choose the ordering of the branch points as $a_1=0$, $a_2=4$, $a_3=(s-2 \sqrt{s}+1)$, $a_4=(s+2 \sqrt{s}+1)$, and we define
\begin{align}
    \Delta(s)=\frac{1}{s (1-s)(9-s)}\,,
\end{align}
which matches the Wronskian up to normalization.
We would like to express our results using the periods
\begin{align}
    \varpi_0 = \oint_A \frac{\dd x}{y}\, ,\qquad \varpi_1 = \oint_B \frac{\dd x}{y}\,.
\end{align}
Note that they differ in normalization to the periods defined in \cite{Broedel:2018qkq}, which are (instead of using the indices $1,2$ we use $0,1$ here)
\begin{align}
    \omega_0=2c_4 \varpi_0\,,\qquad  \omega_1=2c_4 \varpi_1
\end{align}
with $c_4=\frac{1}{2}\sqrt{(a_1-a_3)(a_2-a_4)} = \frac{1}{2} \sqrt{s^2-6 s+8 \sqrt{s}-3}$.
To match our integrand to pure eMPL kernels, we will employ all definitions of \cite{Broedel:2018qkq} in terms of $\omega_{0,1}$, and then we rewrite all periods remaining in the prefactors of the eMPLs in terms of $\varpi_{0,1}$.

Let us start by fixing the first integral. We immediately see that the holomorphic integral $I_0(a,\epsilon)$ is canonical up to a normalization by $\varpi_0$, since
\begin{align}
    I_0(a,0) = \int_C \frac{\mathrm dx}{y} = \frac{\omega_0}{c_4}\int_C \frac{c_4\mathrm dx}{\omega_0\,y} = 2\varpi_0 \int_C\mathrm dx \Psi_0(0,x,\vec a).
\end{align}
We thus choose the first canonical integral as $J_1 = \frac{I_0(a,\epsilon)}{\varpi_0}$.

For the second integral, we again start by considering the derivative of the first canonical integral, $J_1$. Since it has a weight drop due to the double pole, we start with the full $\epsilon$-dependent expression. Using integration by parts, we can rewrite the integrand in terms of the quantities defined above,
\begin{align}\label{eq:J2ellippticSun}
    \partial_s J_1
    &=
    \frac{\Delta}{\varpi_0}\Bigg(\frac{I_0}{\varpi_0} \int_A\dd x \left( \frac{3c_4}{2} \widetilde\Phi_4(s) -\frac{(-3 s^2+22 s-3)}{4y}\right) \nonumber\\
    &+ \int_C \dd x \left( \epsilon  \left(\frac{9c_4}{2} \widetilde \Phi_4(s)+\frac{\left(-7 s^2+14 s+9\right)}{4y} \right)- \frac{3c_4}{2} \widetilde \Phi_4(s)+\frac{1\left(-3 s^2+22 s-3\right)}{4y} \right)u^\epsilon\Bigg) \nonumber\\
    &=
    \frac{\Delta}{\varpi_0}\Bigg(-\frac{3c_4}{2}\int_A\dd x \Phi_4(s) +\epsilon \int_C \dd x   \left(\frac{9c_4}{2} \widetilde \Phi_4(s)+\frac{\left(-7 s^2+14 s+9\right)}{4y}\right)u^\epsilon\Bigg) \, .
\end{align}
First, we can analyze the $\epsilon^0$-piece, setting $u^\epsilon=1$. This contribution vanishes on the $A$-cycle.

As discussed before, while we can use the $B$-cycle to fix the normalization of the second integral, we do not need this information. Let us now consider the next order in $\epsilon$. We rescale the integral by $1/\epsilon$, integrate by parts the 
$1/\epsilon$-term, and we find
\begin{align}\label{eq:integrand Sunrise after IBP}
    \frac{1}{\epsilon}\partial_s J_1 &=    \frac{3c_4\Delta}{2\varpi_0}\int_C \dd x \left(-\frac{2Z_4(x) (-s +x -1 )}{s^2-2 s x-2 s+x^2-2 x+1}-\frac{Z_4(x) }{x-4} + 3\widetilde \Phi_4(x)\right) u^\epsilon\\
   & +\frac{1}{4} \left(-7 s^2+14 s+9\right) \Delta(s)\, J_1\,.
\end{align}
Putting $\epsilon=0$ and rewriting the first line of \cref{eq:integrand Sunrise after IBP} in terms of pure eMPL kernels, we find
\begin{align}
    \frac{1}{\epsilon}\partial_s J_1 &=  \frac{c_4\Delta}{\varpi_0}\int_C \dd x \Bigg( \frac{(s+3)^2 \omega_0 \Psi_0(0,x,\vec a)}{4 c_4^2}-\frac{3 }{2 \omega_0}\Big(6 \Psi_2(\infty,x,\vec a)+i \pi  \Psi_{-1}(4,x,\vec a)\nonumber\\
    &+i \pi  \Psi_{-1}(4,a_3,\vec a)+\pi ^2 \Psi_0(0,x,\vec a)-2 \Psi_{2}(4,x,\vec a)-2 \Psi_{2}(a_3,x,\vec a)-2 \Psi_{2}(a_4,x,\vec a)\Big)\Bigg)\nonumber\\
   & +\frac{1}{4} \left(-7 s^2+14 s+9\right) \Delta(s)\, J_1 \nonumber\\
   &= -\frac{3 }{4}\frac{\Delta}{\varpi_0^2}\int_C \dd x \Bigg(\Big(6 \Psi_2(\infty,x,\vec a)+i \pi  \Psi_{-1}(4,x,\vec a)+i \pi  \Psi_{-1}(4,a_3,\vec a)+\pi ^2 \Psi_0(0,x,\vec a) \nonumber\\
   &-2 \Psi_{2}(4,x,\vec a)-2 \Psi_{2}(a_3,x,\vec a)-2 \Psi_{2}(a_4,x,\vec a)\Big)\Bigg) +\frac{1}{2} \left(-3 s^2+10 s+9\right) \Delta(s)\, J_1 \,,
\end{align}
where we used the special values
\begin{align}
    G_*(\vec a) = \frac{3+s}{6 c_4}\,, \quad Z_4(a_1,\vec a) = Z_4(a_4,\vec a) = 0\,, \quad Z_4(a_2,\vec a) = Z_4(a_3,\vec a) = \frac{i \pi}{\omega_0}.
\end{align}
From here, we see that we can make the integral canonical by multiplying with $\varpi_0^2/\Delta$ and subtracting the extra term proportional to $J_1$, which leads us to the basis
\begin{align}
    J_1 = \frac{I_0(a,\epsilon)}{\varpi_0}\,, \qquad J_2 = \frac{1}{\epsilon}\frac{\varpi_0^2}{\Delta}\partial_s J_1 - \frac{1}{2} \left(9+10 s -3 s^2\right)\varpi_0^2 J_1\,.
\end{align}
It is then easy to verify that this basis indeed satisfies the canonical differential equations
\begin{align}
    \frac\partial{\partial s} \begin{pmatrix}
        J_1 \\ J_2
    \end{pmatrix} = \epsilon
    \begin{pmatrix}
       -\frac{\left(3 s^2-10 s-9\right)  }{2 (1-s) (9-s) s} & \frac{1}{(1-s) (9-s) s \varpi_0^2} \\
 \frac{(3+s)^4 \varpi_0}{4 (1-s) (9-s) s} & -\frac{\left(3 s^2-10 s-9\right)  }{2 (1-s) (9-s) s}
    \end{pmatrix}
    \begin{pmatrix}
        J_1 \\ J_2
    \end{pmatrix}\,.
\end{align}

%---------bibliography if using bibtex
\bibliographystyle{JHEP} % doi-Link, when doi  = {10.1039/C3CC46767H} in .bib file
\bibliography{biblio} % with XeLaTex: needs the same name as main texfile

\providecommand{\href}[2]{#2}\begingroup\raggedright\begin{thebibliography}{10}

\bibitem{Arkani-Hamed:2010pyv}
N.~Arkani-Hamed, J.L.~Bourjaily, F.~Cachazo and J.~Trnka, \emph{{Local
  Integrals for Planar Scattering Amplitudes}},
  \href{https://doi.org/10.1007/JHEP06(2012)125}{\emph{JHEP} {\bfseries 06}
  (2012) 125} [\href{https://arxiv.org/abs/1012.6032}{{\ttfamily 1012.6032}}].

\bibitem{Henn:2013pwa}
J.M.~Henn, \emph{{Multiloop integrals in dimensional regularization made
  simple}}, \href{https://doi.org/10.1103/PhysRevLett.110.251601}{\emph{Phys.
  Rev. Lett.} {\bfseries 110} (2013) 251601}
  [\href{https://arxiv.org/abs/1304.1806}{{\ttfamily 1304.1806}}].

\bibitem{Kummer}
E.E.~Kummer, \emph{{\"{U}ber die Transcendenten, welche aus wiederholten
  Integrationen rationaler Formeln entstehen}}, {\emph{J. reine ang.
  Mathematik} {\bfseries 21} (1840) 74}.

\bibitem{Goncharov:1995}
A.B.~Goncharov, \emph{{Geometry of configurations, polylogarithms, and motivic
  cohomology}}, {\emph{Adv.~Math.} {\bfseries 114} (1995) 197}.

\bibitem{Remiddi:1999ew}
E.~Remiddi and J.A.M.~Vermaseren, \emph{{Harmonic polylogarithms}},
  \href{https://doi.org/10.1142/S0217751X00000367}{\emph{Int. J. Mod. Phys.}
  {\bfseries A15} (2000) 725}
  [\href{https://arxiv.org/abs/hep-ph/9905237}{{\ttfamily hep-ph/9905237}}].

\bibitem{Gehrmann:1999as}
T.~Gehrmann and E.~Remiddi, \emph{{Differential equations for two-loop
  four-point functions}},
  \href{https://doi.org/10.1016/S0550-3213(00)00223-6}{\emph{Nucl. Phys. B}
  {\bfseries 580} (2000) 485}
  [\href{https://arxiv.org/abs/hep-ph/9912329}{{\ttfamily hep-ph/9912329}}].

\bibitem{Goncharov:2010jf}
A.B.~Goncharov, M.~Spradlin, C.~Vergu and A.~Volovich, \emph{{Classical
  Polylogarithms for Amplitudes and Wilson Loops}},
  \href{https://doi.org/10.1103/PhysRevLett.105.151605}{\emph{Phys. Rev. Lett.}
  {\bfseries 105} (2010) 151605}
  [\href{https://arxiv.org/abs/1006.5703}{{\ttfamily 1006.5703}}].

\bibitem{Duhr:2011zq}
C.~Duhr, H.~Gangl and J.R.~Rhodes, \emph{{From polygons and symbols to
  polylogarithmic functions}},
  \href{https://doi.org/10.1007/JHEP10(2012)075}{\emph{JHEP} {\bfseries 1210}
  (2012) 075} [\href{https://arxiv.org/abs/1110.0458}{{\ttfamily 1110.0458}}].

\bibitem{Duhr:2012fh}
C.~Duhr, \emph{{Hopf algebras, coproducts and symbols: an application to Higgs
  boson amplitudes}},
  \href{https://doi.org/10.1007/JHEP08(2012)043}{\emph{JHEP} {\bfseries 08}
  (2012) 043} [\href{https://arxiv.org/abs/1203.0454}{{\ttfamily 1203.0454}}].

\bibitem{tHooft:1972tcz}
G.~'t~Hooft and M.J.G.~Veltman, \emph{{Regularization and Renormalization of
  Gauge Fields}},
  \href{https://doi.org/10.1016/0550-3213(72)90279-9}{\emph{Nucl. Phys. B}
  {\bfseries 44} (1972) 189}.

\bibitem{Bollini:1972ui}
C.G.~Bollini and J.J.~Giambiagi, \emph{{Dimensional Renormalization: The Number
  of Dimensions as a Regularizing Parameter}},
  \href{https://doi.org/10.1007/BF02895558}{\emph{Nuovo Cim. B} {\bfseries 12}
  (1972) 20}.

\bibitem{Kotikov:1990kg}
A.V.~Kotikov, \emph{{Differential equations method: New technique for massive
  Feynman diagrams calculation}},
  \href{https://doi.org/10.1016/0370-2693(91)90413-K}{\emph{Phys. Lett. B}
  {\bfseries 254} (1991) 158}.

\bibitem{Remiddi:1997ny}
E.~Remiddi, \emph{{Differential equations for Feynman graph amplitudes}},
  \href{https://doi.org/10.1007/BF03185566}{\emph{Nuovo Cim. A} {\bfseries 110}
  (1997) 1435} [\href{https://arxiv.org/abs/hep-th/9711188}{{\ttfamily
  hep-th/9711188}}].

\bibitem{Cachazo:2008vp}
F.~Cachazo, \emph{{Sharpening The Leading Singularity}},
  \href{https://arxiv.org/abs/0803.1988}{{\ttfamily 0803.1988}}.

\bibitem{Henn:2020lye}
J.~Henn, B.~Mistlberger, V.A.~Smirnov and P.~Wasser, \emph{{Constructing d-log
  integrands and computing master integrals for three-loop four-particle
  scattering}}, \href{https://doi.org/10.1007/JHEP04(2020)167}{\emph{JHEP}
  {\bfseries 04} (2020) 167}
  [\href{https://arxiv.org/abs/2002.09492}{{\ttfamily 2002.09492}}].

\bibitem{Adams:2016xah}
L.~Adams, C.~Bogner, A.~Schweitzer and S.~Weinzierl, \emph{{The kite integral
  to all orders in terms of elliptic polylogarithms}},
  \href{https://doi.org/10.1063/1.4969060}{\emph{J. Math. Phys.} {\bfseries 57}
  (2016) 122302} [\href{https://arxiv.org/abs/1607.01571}{{\ttfamily
  1607.01571}}].

\bibitem{Adams:2018yfj}
L.~Adams and S.~Weinzierl, \emph{{The $\varepsilon$-form of the differential
  equations for Feynman integrals in the elliptic case}},
  \href{https://doi.org/10.1016/j.physletb.2018.04.002}{\emph{Phys. Lett. B}
  {\bfseries 781} (2018) 270}
  [\href{https://arxiv.org/abs/1802.05020}{{\ttfamily 1802.05020}}].

\bibitem{Dlapa:2022wdu}
C.~Dlapa, J.M.~Henn and F.J.~Wagner, \emph{{An algorithmic approach to finding
  canonical differential equations for elliptic Feynman integrals}},
  \href{https://doi.org/10.1007/JHEP08(2023)120}{\emph{JHEP} {\bfseries 08}
  (2023) 120} [\href{https://arxiv.org/abs/2211.16357}{{\ttfamily
  2211.16357}}].

\bibitem{Yang:2025ofz}
L.L.~Yang and Y.~Zhang, \emph{{From $\mathrm{d} \! \log$ to $\mathrm{d}
  \mathcal{E}$: Canonical Elliptic Integrands and Modular Symbol Letters with
  Pure eMPLs}},  \href{https://arxiv.org/abs/2512.19370}{{\ttfamily
  2512.19370}}.

\bibitem{Chaubey:2025adn}
E.~Chaubey and V.~Sotnikov, \emph{{Elliptic Leading Singularities and Canonical
  Integrands}}, \href{https://doi.org/10.1103/4fjc-lfnx}{\emph{Phys. Rev.
  Lett.} {\bfseries 135} (2025) 101903}
  [\href{https://arxiv.org/abs/2504.20897}{{\ttfamily 2504.20897}}].

\bibitem{Chen:2025hzq}
J.~Chen, L.L.~Yang and Y.~Zhang, \emph{{On an approach to canonicalizing
  elliptic Feynman integrals}},
  \href{https://arxiv.org/abs/2503.23720}{{\ttfamily 2503.23720}}.

\bibitem{Adams:2018bsn}
L.~Adams, E.~Chaubey and S.~Weinzierl, \emph{{Planar Double Box Integral for
  Top Pair Production with a Closed Top Loop to all orders in the Dimensional
  Regularization Parameter}},
  \href{https://doi.org/10.1103/PhysRevLett.121.142001}{\emph{Phys. Rev. Lett.}
  {\bfseries 121} (2018) 142001}
  [\href{https://arxiv.org/abs/1804.11144}{{\ttfamily 1804.11144}}].

\bibitem{Pogel:2022ken}
S.~P\"ogel, X.~Wang and S.~Weinzierl, \emph{{Taming Calabi-Yau Feynman
  Integrals: The Four-Loop Equal-Mass Banana Integral}},
  \href{https://doi.org/10.1103/PhysRevLett.130.101601}{\emph{Phys. Rev. Lett.}
  {\bfseries 130} (2023) 101601}
  [\href{https://arxiv.org/abs/2211.04292}{{\ttfamily 2211.04292}}].

\bibitem{Pogel:2022vat}
S.~P\"ogel, X.~Wang and S.~Weinzierl, \emph{{Bananas of equal mass: any loop,
  any order in the dimensional regularisation parameter}},
  \href{https://doi.org/10.1007/JHEP04(2023)117}{\emph{JHEP} {\bfseries 04}
  (2023) 117} [\href{https://arxiv.org/abs/2212.08908}{{\ttfamily
  2212.08908}}].

\bibitem{Pogel:2022yat}
S.~P\"ogel, X.~Wang and S.~Weinzierl, \emph{{The three-loop equal-mass banana
  integral in \ensuremath{\varepsilon}-factorised form with meromorphic modular
  forms}}, \href{https://doi.org/10.1007/JHEP09(2022)062}{\emph{JHEP}
  {\bfseries 09} (2022) 062}
  [\href{https://arxiv.org/abs/2207.12893}{{\ttfamily 2207.12893}}].

\bibitem{Gorges:2023zgv}
L.~G\"orges, C.~Nega, L.~Tancredi and F.J.~Wagner, \emph{{On a procedure to
  derive \ensuremath{\epsilon}-factorised differential equations beyond
  polylogarithms}}, \href{https://doi.org/10.1007/JHEP07(2023)206}{\emph{JHEP}
  {\bfseries 07} (2023) 206}
  [\href{https://arxiv.org/abs/2305.14090}{{\ttfamily 2305.14090}}].

\bibitem{Duhr:2024uid}
C.~Duhr, F.~Porkert and S.F.~Stawinski, \emph{{Canonical differential equations
  beyond genus one}},
  \href{https://doi.org/10.1007/JHEP02(2025)014}{\emph{JHEP} {\bfseries 02}
  (2025) 014} [\href{https://arxiv.org/abs/2412.02300}{{\ttfamily
  2412.02300}}].

\bibitem{Duhr:2025lbz}
C.~Duhr, S.~Maggio, C.~Nega, B.~Sauer, L.~Tancredi and F.J.~Wagner,
  \emph{{Aspects of canonical differential equations for Calabi-Yau geometries
  and beyond}}, \href{https://doi.org/10.1007/JHEP06(2025)128}{\emph{JHEP}
  {\bfseries 06} (2025) 128}
  [\href{https://arxiv.org/abs/2503.20655}{{\ttfamily 2503.20655}}].

\bibitem{Maggio:2025jel}
S.~Maggio and Y.~Sohnle, \emph{{On canonical differential equations for
  Calabi-Yau multi-scale Feynman integrals}},
  \href{https://doi.org/10.1007/JHEP10(2025)202}{\emph{JHEP} {\bfseries 10}
  (2025) 202} [\href{https://arxiv.org/abs/2504.17757}{{\ttfamily
  2504.17757}}].

\bibitem{e-collaboration:2025frv}
{\scshape {\ensuremath{\varepsilon}}-collaboration} collaboration, \emph{{The
  geometric bookkeeping guide to Feynman integral reduction and
  $\varepsilon$-factorised differential equations}},
  \href{https://arxiv.org/abs/2506.09124}{{\ttfamily 2506.09124}}.

\bibitem{Bree:2025tug}
I.~Bree et~al., \emph{{New algorithms for Feynman integral reduction and
  $\varepsilon$-factorised differential equations}},
  \href{https://arxiv.org/abs/2511.15381}{{\ttfamily 2511.15381}}.

\bibitem{Frellesvig:2021hkr}
H.~Frellesvig, \emph{{On epsilon factorized differential equations for elliptic
  Feynman integrals}},
  \href{https://doi.org/10.1007/JHEP03(2022)079}{\emph{JHEP} {\bfseries 03}
  (2022) 079} [\href{https://arxiv.org/abs/2110.07968}{{\ttfamily
  2110.07968}}].

\bibitem{Frellesvig:2023iwr}
H.~Frellesvig and S.~Weinzierl, \emph{{On $\varepsilon$-factorised bases and
  pure Feynman integrals}},
  \href{https://doi.org/10.21468/SciPostPhys.16.6.150}{\emph{SciPost Phys.}
  {\bfseries 16} (2024) 150}
  [\href{https://arxiv.org/abs/2301.02264}{{\ttfamily 2301.02264}}].

\bibitem{Jiang:2023jmk}
X.~Jiang, X.~Wang, L.L.~Yang and J.~Zhao,
  \emph{{{\ensuremath{\varepsilon}}-factorized differential equations for
  two-loop non-planar triangle Feynman integrals with elliptic curves}},
  \href{https://doi.org/10.1007/JHEP09(2023)187}{\emph{JHEP} {\bfseries 09}
  (2023) 187} [\href{https://arxiv.org/abs/2305.13951}{{\ttfamily
  2305.13951}}].

\bibitem{Giroux:2024yxu}
M.~Giroux, A.~Pokraka, F.~Porkert and Y.~Sohnle, \emph{{The soaring kite: a
  tale of two punctured tori}},
  \href{https://doi.org/10.1007/JHEP05(2024)239}{\emph{JHEP} {\bfseries 05}
  (2024) 239} [\href{https://arxiv.org/abs/2401.14307}{{\ttfamily
  2401.14307}}].

\bibitem{Marzucca:2025eak}
R.~Marzucca, A.J.~McLeod and C.~Nega, \emph{{Two-Loop Master Integrals for
  Mixed QCD-EW Corrections to $gg \to H$ Through $\mathcal{O}(\epsilon^2)$}},
  \href{https://arxiv.org/abs/2501.14435}{{\ttfamily 2501.14435}}.

\bibitem{Becchetti:2025rrz}
M.~Becchetti, F.~Coro, C.~Nega, L.~Tancredi and F.J.~Wagner, \emph{{Analytic
  two-loop amplitudes for $ q\overline{q}\to \gamma \gamma $ and gg
  {\textrightarrow} {\ensuremath{\gamma}}{\ensuremath{\gamma}} mediated by a
  heavy-quark loop}},
  \href{https://doi.org/10.1007/JHEP06(2025)033}{\emph{JHEP} {\bfseries 06}
  (2025) 033} [\href{https://arxiv.org/abs/2502.00118}{{\ttfamily
  2502.00118}}].

\bibitem{Becchetti:2025oyb}
M.~Becchetti, C.~Dlapa and S.~Zoia, \emph{{Canonical differential equations for
  the elliptic two-loop five-point integral family relevant to $t\bar t +$jet
  production at leading colour}},
  \href{https://arxiv.org/abs/2503.03603}{{\ttfamily 2503.03603}}.

\bibitem{Coro:2025vgn}
F.~Coro, C.~Nega, L.~Tancredi and F.J.~Wagner, \emph{{Analytic two-loop
  amplitudes for di-jet and {\ensuremath{\gamma}}+jet production mediated by a
  heavy-quark loop}},
  \href{https://doi.org/10.1007/JHEP01(2026)090}{\emph{JHEP} {\bfseries 01}
  (2026) 090} [\href{https://arxiv.org/abs/2509.15315}{{\ttfamily
  2509.15315}}].

\bibitem{Duhr:2022pch}
C.~Duhr, A.~Klemm, F.~Loebbert, C.~Nega and F.~Porkert,
  \emph{{Yangian-Invariant Fishnet Integrals in Two Dimensions as Volumes of
  Calabi-Yau Varieties}},
  \href{https://doi.org/10.1103/PhysRevLett.130.041602}{\emph{Phys. Rev. Lett.}
  {\bfseries 130} (2023) 041602}
  [\href{https://arxiv.org/abs/2209.05291}{{\ttfamily 2209.05291}}].

\bibitem{Duhr:2023eld}
C.~Duhr, A.~Klemm, F.~Loebbert, C.~Nega and F.~Porkert, \emph{{The Basso-Dixon
  formula and Calabi-Yau geometry}},
  \href{https://doi.org/10.1007/JHEP03(2024)177}{\emph{JHEP} {\bfseries 03}
  (2024) 177} [\href{https://arxiv.org/abs/2310.08625}{{\ttfamily
  2310.08625}}].

\bibitem{Duhr:2024hjf}
C.~Duhr, A.~Klemm, F.~Loebbert, C.~Nega and F.~Porkert, \emph{{Geometry from
  integrability: multi-leg fishnet integrals in two dimensions}},
  \href{https://doi.org/10.1007/JHEP07(2024)008}{\emph{JHEP} {\bfseries 07}
  (2024) 008} [\href{https://arxiv.org/abs/2402.19034}{{\ttfamily
  2402.19034}}].

\bibitem{Duhr:2024bzt}
C.~Duhr, F.~Gasparotto, C.~Nega, L.~Tancredi and S.~Weinzierl, \emph{{On the
  electron self-energy to three loops in QED}},
  \href{https://doi.org/10.1007/JHEP11(2024)020}{\emph{JHEP} {\bfseries 11}
  (2024) 020} [\href{https://arxiv.org/abs/2408.05154}{{\ttfamily
  2408.05154}}].

\bibitem{Forner:2024ojj}
F.~Forner, C.~Nega and L.~Tancredi, \emph{{On the photon self-energy to three
  loops in QED}},  \href{https://arxiv.org/abs/2411.19042}{{\ttfamily
  2411.19042}}.

\bibitem{Duhr:2025kkq}
C.~Duhr, S.~Maggio, F.~Porkert, C.~Semper and S.F.~Stawinski, \emph{{Three-loop
  banana integrals with four unequal masses}},
  \href{https://doi.org/10.1007/JHEP12(2025)034}{\emph{JHEP} {\bfseries 12}
  (2025) 034} [\href{https://arxiv.org/abs/2507.23061}{{\ttfamily
  2507.23061}}].

\bibitem{Duhr:2025ouy}
C.~Duhr and S.~Maggio, \emph{{Three-loop banana integrals with three equal
  masses}},  \href{https://arxiv.org/abs/2511.19245}{{\ttfamily 2511.19245}}.

\bibitem{Pogel:2025bca}
S.~P{\"o}gel, T.~Teschke, X.~Wang and S.~Weinzierl, \emph{{The unequal-mass
  three-loop banana integral}},
  \href{https://doi.org/10.1007/JHEP01(2026)021}{\emph{JHEP} {\bfseries 01}
  (2026) 021} [\href{https://arxiv.org/abs/2507.23594}{{\ttfamily
  2507.23594}}].

\bibitem{Bern:2025wyd}
Z.~Bern, E.~Herrmann, R.~Roiban, M.S.~Ruf, A.V.~Smirnov, S.~Smith et~al.,
  \emph{{Scattering Amplitudes and Conservative Binary Dynamics at $O(G^5)$
  without Self-Force Truncation}},
  \href{https://arxiv.org/abs/2512.23654}{{\ttfamily 2512.23654}}.

\bibitem{Klemm:2024wtd}
A.~Klemm, C.~Nega, B.~Sauer and J.~Plefka, \emph{{Calabi-Yau periods for black
  hole scattering in classical general relativity}},
  \href{https://doi.org/10.1103/PhysRevD.109.124046}{\emph{Phys. Rev. D}
  {\bfseries 109} (2024) 124046}
  [\href{https://arxiv.org/abs/2401.07899}{{\ttfamily 2401.07899}}].

\bibitem{Driesse:2024feo}
M.~Driesse, G.U.~Jakobsen, A.~Klemm, G.~Mogull, C.~Nega, J.~Plefka et~al.,
  \emph{{High-precision black hole scattering with Calabi-Yau manifolds}},
  \href{https://arxiv.org/abs/2411.11846}{{\ttfamily 2411.11846}}.

\bibitem{Driesse:2026qiz}
M.~Driesse, G.U.~Jakobsen, G.~Mogull, C.~Nega, J.~Plefka, B.~Sauer et~al.,
  \emph{{Conservative Black Hole Scattering at Fifth Post-Minkowskian and
  Second Self-Force Order}},
  \href{https://arxiv.org/abs/2601.16256}{{\ttfamily 2601.16256}}.

\bibitem{Bern:2026oqp}
Z.~Bern, A.~Jackman, G.~Mansfield and M.S.~Ruf, \emph{{Classical Gravitational
  Scattering from the Ultraviolet and the Absence of Calabi-Yau Integrals in
  the Conservative Sector at $O(G^5)$}},
  \href{https://arxiv.org/abs/2603.15383}{{\ttfamily 2603.15383}}.

\bibitem{Bourjaily:2020hjv}
J.L.~Bourjaily, N.~Kalyanapuram, C.~Langer, K.~Patatoukos and M.~Spradlin,
  \emph{{Elliptic, Yangian-Invariant {\textquotedblleft}Leading
  Singularity{\textquotedblright}}},
  \href{https://doi.org/10.1103/PhysRevLett.126.201601}{\emph{Phys. Rev. Lett.}
  {\bfseries 126} (2021) 201601}
  [\href{https://arxiv.org/abs/2012.14438}{{\ttfamily 2012.14438}}].

\bibitem{Bourjaily:2021vyj}
J.L.~Bourjaily, N.~Kalyanapuram, C.~Langer and K.~Patatoukos,
  \emph{{Prescriptive unitarity with elliptic leading singularities}},
  \href{https://doi.org/10.1103/PhysRevD.104.125009}{\emph{Phys. Rev. D}
  {\bfseries 104} (2021) 125009}
  [\href{https://arxiv.org/abs/2102.02210}{{\ttfamily 2102.02210}}].

\bibitem{Duhr:2024xsy}
C.~Duhr, F.~Porkert, C.~Semper and S.F.~Stawinski, \emph{{Self-duality from
  twisted cohomology}},
  \href{https://doi.org/10.1007/JHEP03(2025)053}{\emph{JHEP} {\bfseries 03}
  (2025) 053} [\href{https://arxiv.org/abs/2408.04904}{{\ttfamily
  2408.04904}}].

\bibitem{Duhr:2025xyy}
C.~Duhr, S.~Maggio, F.~Porkert, C.~Semper, Y.~Sohnle and S.F.~Stawinski,
  \emph{{Canonical differential equations and intersection matrices}},
  \href{https://doi.org/10.1007/JHEP02(2026)211}{\emph{JHEP} {\bfseries 02}
  (2026) 211} [\href{https://arxiv.org/abs/2509.17787}{{\ttfamily
  2509.17787}}].

\bibitem{Brown:2011wfj}
F.C.S.~Brown and A.~Levin, \emph{{Multiple Elliptic Polylogarithms}},
  \href{https://arxiv.org/abs/1110.6917}{{\ttfamily 1110.6917}}.

\bibitem{Mastrolia:2018uzb}
P.~Mastrolia and S.~Mizera, \emph{{Feynman Integrals and Intersection Theory}},
  \href{https://doi.org/10.1007/JHEP02(2019)139}{\emph{JHEP} {\bfseries 02}
  (2019) 139} [\href{https://arxiv.org/abs/1810.03818}{{\ttfamily
  1810.03818}}].

\bibitem{Mizera:2017rqa}
S.~Mizera, \emph{{Scattering Amplitudes from Intersection Theory}},
  \href{https://doi.org/10.1103/PhysRevLett.120.141602}{\emph{Phys. Rev. Lett.}
  {\bfseries 120} (2018) 141602}
  [\href{https://arxiv.org/abs/1711.00469}{{\ttfamily 1711.00469}}].

\bibitem{Bargiela:2025vwl}
P.~Bargiela, H.~Frellesvig, R.~Marzucca, R.~Morales, F.~Seefeld, M.~Wilhelm
  et~al., \emph{{The spectrum of Feynman-integral geometries at two loops}},
  \href{https://arxiv.org/abs/2512.13794}{{\ttfamily 2512.13794}}.

\bibitem{GriffithsHarris}
P.~Griffiths and J.~Harris, \emph{Principles of Algebraic Geometry}, Wiley
  (1978).

\bibitem{Vanhove:2014wqa}
P.~Vanhove, \emph{{The physics and the mixed Hodge structure of Feynman
  integrals}}, \href{https://doi.org/10.1090/pspum/088/01455}{\emph{Proc. Symp.
  Pure Math.} {\bfseries 88} (2014) 161}
  [\href{https://arxiv.org/abs/1401.6438}{{\ttfamily 1401.6438}}].

\bibitem{Bonisch:2021yfw}
K.~B\"onisch, C.~Duhr, F.~Fischbach, A.~Klemm and C.~Nega, \emph{{Feynman
  integrals in dimensional regularization and extensions of Calabi-Yau
  motives}}, \href{https://doi.org/10.1007/JHEP09(2022)156}{\emph{JHEP}
  {\bfseries 09} (2022) 156}
  [\href{https://arxiv.org/abs/2108.05310}{{\ttfamily 2108.05310}}].

\bibitem{ChenSymbol}
K.T.~Chen, \emph{{Iterated path integrals}}, {\emph{Bull.\ Amer.\ Math.\ Soc.}
  {\bfseries 83} (1977) 831}.

\bibitem{Broedel:2018qkq}
J.~Broedel, C.~Duhr, F.~Dulat, B.~Penante and L.~Tancredi, \emph{{Elliptic
  Feynman integrals and pure functions}},
  \href{https://doi.org/10.1007/JHEP01(2019)023}{\emph{JHEP} {\bfseries 01}
  (2019) 023} [\href{https://arxiv.org/abs/1809.10698}{{\ttfamily
  1809.10698}}].

\bibitem{Primo:2016ebd}
A.~Primo and L.~Tancredi, \emph{{On the maximal cut of Feynman integrals and
  the solution of their differential equations}},
  \href{https://doi.org/10.1016/j.nuclphysb.2016.12.021}{\emph{Nucl. Phys. B}
  {\bfseries 916} (2017) 94}
  [\href{https://arxiv.org/abs/1610.08397}{{\ttfamily 1610.08397}}].

\bibitem{Primo:2017ipr}
A.~Primo and L.~Tancredi, \emph{{Maximal cuts and differential equations for
  Feynman integrals. An application to the three-loop massive banana graph}},
  \href{https://doi.org/10.1016/j.nuclphysb.2017.05.018}{\emph{Nucl. Phys. B}
  {\bfseries 921} (2017) 316}
  [\href{https://arxiv.org/abs/1704.05465}{{\ttfamily 1704.05465}}].

\bibitem{Frellesvig:2017aai}
H.~Frellesvig and C.G.~Papadopoulos, \emph{{Cuts of Feynman Integrals in Baikov
  representation}}, \href{https://doi.org/10.1007/JHEP04(2017)083}{\emph{JHEP}
  {\bfseries 04} (2017) 083}
  [\href{https://arxiv.org/abs/1701.07356}{{\ttfamily 1701.07356}}].

\bibitem{Bosma:2017ens}
J.~Bosma, M.~Sogaard and Y.~Zhang, \emph{{Maximal Cuts in Arbitrary
  Dimension}}, \href{https://doi.org/10.1007/JHEP08(2017)051}{\emph{JHEP}
  {\bfseries 08} (2017) 051}
  [\href{https://arxiv.org/abs/1704.04255}{{\ttfamily 1704.04255}}].

\bibitem{Broedel:2017kkb}
J.~Broedel, C.~Duhr, F.~Dulat and L.~Tancredi, \emph{{Elliptic polylogarithms
  and iterated integrals on elliptic curves. Part I: general formalism}},
  \href{https://doi.org/10.1007/JHEP05(2018)093}{\emph{JHEP} {\bfseries 05}
  (2018) 093} [\href{https://arxiv.org/abs/1712.07089}{{\ttfamily
  1712.07089}}].

\bibitem{Frellesvig:2024ymq}
H.~Frellesvig, \emph{{The Loop-by-Loop Baikov Representation -- Strategies and
  Implementation}},  \href{https://arxiv.org/abs/2412.01804}{{\ttfamily
  2412.01804}}.

\bibitem{Studerus:2009ye}
C.~Studerus, \emph{{Reduze~{\textendash} Feynman integral reduction in C++}},
  \href{https://doi.org/10.1016/j.cpc.2010.03.012}{\emph{Comput. Phys. Commun.}
  {\bfseries 181} (2010) 1293}
  [\href{https://arxiv.org/abs/0912.2546}{{\ttfamily 0912.2546}}].

\bibitem{vonManteuffel:2012je}
A.~von Manteuffel and C.~Studerus, \emph{{Top quark pairs at two loops and
  Reduze 2}}, \href{https://doi.org/10.22323/1.151.0059}{\emph{PoS} {\bfseries
  LL2012} (2012) 059} [\href{https://arxiv.org/abs/1210.1436}{{\ttfamily
  1210.1436}}].

\end{thebibliography}\endgroup

\end{document}